# THE INTERPLAY BETWEEN CHEMISTRY AND MECHANICS IN THE TRANSDUCTION OF A MECHANICAL SIGNAL INTO A BIOCHEMICAL FUNCTION

Francesco Valle[a], Massimo Sandal[a], Bruno Samorì[a*]

[a]University of Bologna, G.Moruzzi Department of Biochemistry , Via Irnerio 48, 40126 Bologna, Italy

* Corresponding author.
*E-mail addresses:* francesco.valle@unibo.it (F.Valle), massimo.sandal@unibo.it (M.Sandal), samori@alma.unibo.it  (B.Samorì)

**Abstract:**
There are many processes in biology in which mechanical forces are generated. Force-bearing networks can transduce locally developed mechanical signals very extensively over different parts of the cell or tissues. In this article we conduct an overview of this kind of mechanical transduction, focusing in particular on the multiple layers of complexity displayed by the mechanisms that control and trigger the conversion of a mechanical signal into a biochemical function. Single molecule methodologies, through their capability to introduce the force in studies of biological processes in which mechanical stresses are developed, are unveiling subtle intertwining mechanisms between chemistry and mechanics and in particular are revealing how chemistry can control mechanics. The possibility that chemistry interplays with mechanics should be always considered in biochemical studies.

Table of Contents:

# INTRODUCTION

In the middle of the XIXth century, the model of the cell according to classical biochemistry was that of a chemical reactor containing a myriad of chemical species undergoing a multitude of reactions brought about by diffusions and random collisions. This picture has been drastically changed in the last fifty years by the discovery that many of those reactions involve a mechanical force leading to a directional movement and transport of the chemical species. In most cases this mechanical force is generated by complex molecular structures behaving in a machine-like fashion [1]. A true world of forces acting inside the cell has gradually been disclosed. On the cell surface also a myriad of tensions are generated by the motility and proliferation of the surrounding cells, by osmotic forces [2], and in some cases also by the pressure of the circulating fluids in the body. Cells are also able to respond to the stiffness of their microenvironment, as most tissue cells do through their adhesion [3].

The overall cell shape results from the development over time of this world of forces. The cell shape has been proved to be an important modulator of cellular physiology and function in a variety of tissues [4] and in their repairing [5], in connective tissue homeostasis [6], in skeletal muscle gene regulation and hypertrophy [7], among others. Furthermore, as indicated by recent micro-array studies, changes in cell shape also affect gene expression of proteins of the cytoskeleton, proliferation, transcription, translation, Extra Cellular Matrix (ECM) production and inter- and intracellular signalling complexes [8].

Cells are therefore able to sense and read external forces quantitatively, to transmit them from outside to inside [8, 9], and also to respond to them mechanically through feedback loops acting in the opposite direction [3].

The cell signalling pathways that make this kind of mechanical conversation between the outside and the inside of the cell possible require not only transmission of the mechanical stresses but also their conversion into a distinct set of biochemical functions through specific binding to affinity or catalytic sites.

In this review we start from the mechanisms of the conversion of a stress into a biochemical signal through the formation of a bond. In the first section we look at the crucial role played in this context by the interplay between the kinetics of formation and the lifetime of this bond under an external force. We will see how this interplay is controlled by the mechanical deformation of the energy landscape of the bond.

The second section tackles the problem of the mechanical activation of the same bond within complex protein networks, like those transmitting and driving cellular mechanical stresses. The mechanisms of transmission of forces that can be brought about by protein unfolding-refolding cycles under different force waveforms are overviewed. In this context, a few very enlightening experiments recently reported are discussed, like those concerning the unfoldase activity required for protein translocation through a mitocondria pore, those on the chaperone-assisted refolding activity, and those on the mechanical resistance of the natural adhesives present in several mineralized parts of living organisms.

The third section moves into a higher degree of complexity - into the chemistry that can control protein mechanics. More specifically, we will see how, according to a mechanism recently proposed by us, redox equilibria can tune the local delivery of the mechanical stresses onto a specific binding site and, in this way, their conversion into biochemical signals.

## 1. THE CONVERSION OF A MECHANICAL STRESS INTO A BIOCHEMICAL SIGNAL THROUGH BINDING EVENTS

The control of cellular morphology and functions through mechanical signals requires the conversion of the locally delivered forces into biochemical functions. The triggering of these functions must involve the specific binding to affinity or catalytic sites. Sites of this kind have been identified in the proteins of the extra- and intracellular force-bearing networks, as in the case of fibronectin (Fn), a cell adhesion protein abundant in serum and in many matrices, that is constituted by more than 50 modular repeats. Many molecular recognition sites for other matrix proteins and integrins have been identified in loop regions of the beta sandwich motifs of this protein. In addition to them, cryptic binding sites buried in the folded structures of its FnIII modules have been also recognized (see [9]).

Mechanical stresses can drive the conformational changes required for the exposure of cryptic sites, thus making their binding possible [10, 11] (see Figure 1). We can expect that by this mechanism phosphorylation sites can be exposed at the right moment, enzymatically active sites can be activated through their straining and deformation, physiologically-significant intermediate states can be created and stabilized, etc.

Traditionally a single bond or a complex adhesion event that is sustained by a set of molecular interactions is described in terms of the association kinetics of their components. This description is commonly based on measurements of binding affinities performed under reversible equilibrium.

Also the dissociation and the lifetime of the same binding event can play a crucial role in the control of signal switching mechanisms. A bond that is formed in 1 sec and breaks after 1 min has the same affinity as the bond that requires 1 min to form (maybe because of a complex recognition process) and 1 hour to dissociate. It might be that not the faster, but the slower binding can trigger a biochemical signal, just because of its longer lifetime.

The behaviour of a mechanically activated switch must therefore be described in terms of the relation between force, lifetime and affinity, i.e. in terms of the dynamics of formation and dissociation of the interactions upon which it relies. This dynamics is controlled by the mechanical deformation of the energy landscape of the same interactions. We will see in section 2 that the energy landscape of a binding event that triggers a distinct biochemical signal must be appropriately integrated with the energy landscape that controls the mechanical activation of the relative binding sites. Single molecule methodologies now make it possible to map both types of energy landscapes under an external force. The relations between force, lifetime and affinity of a chemical interaction and its integration in the energy landscape of a protein network can now be tackled by bringing these methodologies into play.

## 1.1 The dynamics under stress of a binding interaction having a single-barrier energy landscape

As sketched in Figure 2, a bond can be described by an energy profile where a barrier confines the bound state away from the dissociated one. The energy difference between the bound and the dissociated state controls the relative populations of the two states. This energy gap can be measured by traditional bulk ensemble techniques, but as pointed out above, by measuring only the energy difference we learn very little about the relations between force, lifetime and affinity that describe the dynamics of a binding event.

In general, any bond or interaction has a limited lifetime, so if we wait for a sufficient amount of time we will observe it to break. Whenever an external force $f$ that breaks apart the bound components is applied, the bond lifetime under that force is described by equation 1,

$$t_{off}(f) = \frac{1}{k_{off}(f)} = t_D \exp\left[\frac{E_b(f)}{k_B T}\right] \qquad (1)$$

where $t_D$ is the diffusive relaxation time and $E_b(f)$ is the height of the energy barrier that confines the bound from the dissociated state.

The pulling force tilts the energy landscape (Figure 2B). Evans and Ritchie [12] were the first to model the change induced by an externally applied force on the energy profile of a bond, starting from a pioneering study by Bell [13], and on this basis they analysed in detail the physical quantities that, as we will see below, can be measured in force spectroscopy experiments. Later on, a number of works implemented and refined this approach [14-18].

The energy barrier at the transition state is reduced by the application of the external force, the dissociated state is made more favoured, and therefore the lifetime of the bond decreases. The application of an external force properly directed thus supplies some encouragement to the binding to fail. The energy distortion shown in Figure 2B quantitatively changes the bond lifetime in an exponential way, as described by equation 2, where the height of the barrier is reduced of an amount corresponding to the work performed along the reaction coordinate by the force $f$ along a distance $x_b$ (the distance of the barrier from the minimum).

$$t_{off}(f) = \frac{1}{k_{off}(f)} = t_D \exp\left[\frac{E_b(f)}{k_B T}\right] = t_D \exp\left[\frac{E_b(0) - fx_b}{k_B T}\right] = t_{off}(0) \exp\left[-\frac{fx_b}{k_B T}\right] \qquad (2)$$

The overall effect is a downward tilting of the energy landscape. The extent of the reduction of the height of the barrier is greater the further the location $x_b$ from the equilibrium configuration (Figure 2C).

To relate the bond lifetime to the measured values of the force at which a bond fails, one has to consider the dissociation equation of an isolated pair of interacting molecules. The master equation is

$$\frac{dp_1}{dt} = -k_{off}(t) p_1(t) + k_{on}(t) p_0(t) \qquad (3)$$

where $p_1(t)$ is is the likelihood of being in the bound state and $p_0(t)=1- p_1(t)$ is the likelihood of being dissociated. $k_{off}$ and $k_{on}$ are the dissociation and binding constants of the complex.

This equation, when a force is applied as to separate the bound components, reduces to

$$\frac{dp_1}{dt} = -k_{off}(t) p_1(t) \qquad (4)$$

because, keeping on pulling, they move apart and thus the rebinding probability vanishes.

When an elastic probe (AFM cantilever, optical trap, biomembrane force probe), acts as a pulling spring that moves apart, at a constant speed, the two bound moieties (velocity-clamp mode), the applied force increases with time. The relationship between dynamics and mechanics is better described by the loading rate $r_f(f) = \frac{df}{dt} = k_s(f) v_s$, where $k_s$ is the spring constant of the full system composed of the force sensor and the tethered molecules and $v_s$ is its pulling speed. By using this parameter the force becomes the integration variable, instead of the time. By introducing the force dependent expression of $k_{off}(f)$ in the master equation and solving the differential equation thus obtained, the probability density of the bond survival as a function of the applied force, i.e. the distribution of the unbinding forces, is obtained.

$$\omega(f) = \frac{dp_1(f)}{df} = \frac{k_{off}(0)}{r_f} \exp\left[\frac{fx_b}{k_B T} + \frac{k_{off}(0)}{r_f} \frac{k_B T}{x_b}\left(1 - \exp\left(\frac{fx_b}{k_B T}\right)\right)\right] \qquad (5)$$

As a consequence, it is possible to calculate the value of the force at which this distribution has a maximum, corresponding to the most probable rupture force measured in the single molecule force spectroscopy (SMFS) experiments.

$$f(r) = \omega(f) = \frac{k_B T}{x_b} \ln\left(\frac{r_f}{k_{off}(0)} \frac{x_b}{k_B T}\right) \qquad (6)$$

**1.1.1 The lifetime of a binding event and the location of the energy barrier by which it is confined**

The most probable rupture force ($f^*$) of the interaction depends upon the loading rate ($r_f$) because the likelihood of the bond failing under an external force depends on the time spent at the same force. Under the application of a ramp of force the time spent by the bond at a certain force is controlled by the pulling velocity. If the velocity is so high that this time is shorter than the bond lifetime at the same force, the bond is likely to fail only at a greater force. The value of the most probable rupture force therefore increases with the pulling velocity, i.e. with the loading rate.

This dependence is expected to be a weaker one, like the logarithmic one, (see Equation 6) because it comes out from the balance of two effects. On pulling faster, the most probable rupture force becomes greater, but at the same time, the dominant barrier is lowered, thus leading to a decrease of the bond lifetime and also of the most probable force at which the rupture takes place.

The force-induced reduction of the height of the barrier being greater the further is its location along the reaction coordinate (Figure 2C), as predicted by Equation 6, a linear plot of the most-probable-rupture-force *versus* the logarithm of the loading rate is therefore obtained (see Figure 2D). This dependence makes it possible to determine not only the barrier location ($x_b$), but also the unbinding constant at zero force ($k_{off}(0)$).

The spatial location of the barrier ($x_b$) is given by the slope of that linear relationship, and therefore if $k_s$ is assumed to be constant, only knowledge of $v$ is needed. On the other hand $k_{off}(0)$ depends both upon the slope and the constant term. This is the reason why an accurate evaluation of the effective loading rate $k_s v_s$ is needed. The velocity $v$ is set within the dynamic range accessible to the instrument and is known with very high accuracy. The value of $k_s$ was assumed by many authors to be determined by the spring constant of the cantilever only. Most recently it has become understood that $k_s$ should include also the elasticity of the stretched molecule. On the other hand, the elasticity of the molecule depends on the conformational space it can explore. This space is gradually narrowed on increasing the extent of the stress applied to the molecule. This is the reason why Dettmann *et al.* [19] proposed to evaluate $k_s$ values from the slope of the final part of the force peaks, i.e. when the molecule is in a stretched state that is very close to the bond rupture point. Unfortunately, in most cases, noise and irregularities of the peak profile often make somewhat arbitrary the definition of a meaningful 'slope'.

This problem can be circumvented by obtaining $k_{off}$ and $x_b$ from Monte Carlo simulations of the mechanical unfolding. The simulations normally assume a two-state model for the unfolding potential barrier and the worm-like chain (WLC) elasticity model (see section 2.1.1) for protein elongation. The energy parameters of the barrier are adjusted to fit the experimental force distribution and the dependence of the rupture force data from the pulling speed. With this method the loading rate is implicit in the calculation, being a function of cantilever elastic constant and of molecule persistence length: these parameters must be given to the simulation as explicit data [20]. If there is some knowledge about the interactions involved in mechanical unfolding, more refined models can be used [21].

Evaluations of not only the intrinsic rate coefficient and location of transition state, but also of the free energy of activation, can be extracted by taking into account that the tilting of the energy landscape not only decreases the height of the barriers, but also subtly shifts them towards the equilibrium minima [22]. Because of this shift, the mean rupture force becomes a non linear function of the logarithm of the loading rate. Nevertheless at low to intermediate speeds a linear fit of this model is quite good. Dudko *et al.* [22] therefore warn that, in this case, one might incorrectly attribute the curvature appearing at higher force loading rates to the switch from one dominant free-energy barrier to another.

## 1.2 The dynamics of an adhesion interaction with a complex energy landscape.

In biological environments where forces are developed, the adhesions are normally sustained by a set of non-covalent interactions that act in concert. The resulting energy landscapes of these adhesions are much more complex than those considered in the previous paragraph. Furthermore dynamic systems are characterized by nearly degenerate and shallow energy minima in their energy landscapes. High barriers are in fact not compatible with dynamic mechanisms of signal triggering in response to mechanical signals. High barriers lead to static systems unable to switch between different pathways. In an energy landscape with nearly degenerate and shallow energy minima, instead, the thermal bath ensures the fluctuations for the proofreading of the different paths [23]. On the other hand, forces externally applied can impose directionality to this proofreading by deforming the energy landscapes so as to drive the system towards the requested function. Therefore, even more in the case of this type of complex and dynamic adhesions, we need to explore their energy landscapes under an applied force.

**1.2.1 Mapping the energy landscape of a complex adhesion event**

If more than one barrier is present in the energy landscape (Figure 3A), different linear regimes come out in sequence in the plots of the most-probable-rupture-force *versus* the logarithm of the loading rate (Figure 3B). A staircase of different rates constant under an increasing loading force is obtained: each linear regime corresponds to the overcoming of a single energy barrier along the unbinding pathway [24, 25]. It must be remembered that the abovementioned analysis by Szabo *et* al. [22] warns about the possibiity of deviation from this kind of linearity. In principle, by varying the force-loading rate on the molecule, one can make different barriers emerge as the rate-determining ones. Also an outer barrier that is sufficiently high to control the kinetics of the thermal unbinding, is in turn overwhelmed by a smaller inner barrier that is made to emerge so much as to become the rate-determining in the mechanical experiment. Contrary to experiments in thermodynamic equilibrium, inner barriers can thus be revealed.

The dissociation energy landscape can therefore be mapped. For each regime corresponding to the different barriers that in sequence become dominant, the dissociation rate ($k_{off}(0)$) and the corresponding barrier width ($x_b$) are evaluated, as shown in the case of one only barrier (see section 1.1.1 )[24-28].

If the energy landscape is much more complex than that depicted in Figure 3, due to the present experimental limitations, its mapping will loose most of its details - it might be characterized only in terms of its main energy barriers along the unfolding pathways [29].

**1.2.2 Tailoring the energy landscape for adhesion events that must be either transient or firm under a mechanical stress**

The capability of mapping energy landscapes allows us learn how it is possible to modulate not only the rupture force and the lifetime of an interaction, but also the stability of an intermediate that in the absence of an external force would not be favoured. Therefore on this basis we can also learn how a bond dissociation can switch between two different pathways just by changing the kind of force ramp [30].

We can tailor experiments to address questions like: How have different cell-protein adhesions been optimized by nature to ensure a specific mechanical performance under a hydrodynamic flow?  For instance, what is the shape of the energy landscape that drives the control of the rolling adhesion of leukocytes in the blood vessels? Or, what is the shape of the energy landscape of the interactions mediated by fibronectin that sustain the infections of staphylococci to the endothelial cells lining the vessel?

The two energy landscapes should be drastically different. In fact, in the former case, the velocity of the leukocytes must be kept constant and therefore the binding must be more effective at high hydrodynamic stress in order to slow them down [30-32].  In the latter case, at high hydrodynamic stress the binding must be less effective - in fact no infections at blood vessel entrances and bifurcations (turbulent flow, high shear stress) are normally observed [31].

Leukocytes roll along the endothelium of post-capillary venules in response to inflammatory signals. Rolling under the hydrodynamic drag forces of blood flow is mediated by the L-selectin/P-selectin glycoprotein ligand 1 (PSGL-1) interaction (see[31] and references therein). As shown in Figure 3B, in the loading rate range accessible to the AFM-based SMFS methodology, a two-barriers energy landscape that control the L-selectin/PSGL-1 adhesion was mapped [32]. These barriers are located at 0.6 and 4.0 A° from the equilibrium bound position and the lifetimes extrapolated at zero force, t(0), of the corresponding two minima are 0. 01s and 0.33s. In general, barriers characterized by short $x_b$ can sustain high forces but small deformations and thus display a 'brittle' character; conversely, barriers possessing large $x_b$ are 'elastic' or compliant, being capable of sustaining low forces but larger deformations [33]. In the case of the L-selectin/PSGL-1 adhesion (Figure 3B), the inner barrier at 0.6 A°, being so close to the equilibrium minimum, can provide, for very short loading times, the high strength attachment needed to initiate the leukocytes tethering and to interrupt cell translocation in the flow. Whenever this very brittle interaction has failed, the adhesion falls into the second pocket where an additional interaction is brought into play. This interaction because of the large width of the corresponding pocket, is very compliant and therefore capable of ensuring longer residence at vessels but at the same time has a reduced capability to sustain forces (Figure 2C). This latter request is less stringent when the cells have been tethered already, because a smaller stress is exerted on the adhesion.

The adhesion between Fibronectin (Fn) and its adhesins mediates the bacterial invasions and persistent infections of *Staphylococcus epidermis*. Also this adhesion takes place under physiological shear rates. Its mechanical performance must be drastically different from that of L-selectin/PSGL-1 adhesion. Because of the limited range of the spanned loading rates, only a single linear regime in the force spectrum of the Fn/Fn–adhesin was found, corresponding to the pocket sketched in Figure 3E. On the other hand, a comparison of the time survival of the bond extrapolated to zero force estimated by force spectroscopy experiments with the bond lifetime measured by Surface Plasmon Resonance for the thermally activated dissociation, demonstrated that the energy landscape of Fn/Fn–adhesin interaction is controlled by more than one potential barrier and the regime so far identified by force spectroscopy must correspond to an inner barrier [31]. The corresponding values of $x_b$ and $k_{off}(0)$ are typical of an elastic interaction selected to sustain relatively large deformations at low forces and, at the same time, low loading rates for significantly longer times. This kind of performance fits the strong liability of bacterial adhesion to high, turbulent flow. On this basis we can understand the physical basis of the very unlikely location of the relative infections at blood vessel entrances and bifurcations, where the flow is turbulent, and stresses are developed [31].

The capability of mapping energy landscapes makes it possible also to answer questions like: how the energy landscape of a rolling adhesion, like that in Figure 3C can be transformed into a firmer adhesion? One might propose to reduce the brittleness of the interactions sustaining the first step of the rolling adhesion by enlarging the first pocket. The same pocket might be also deepened so as to stabilize the corresponding bound state. Furthermore, the second pocket might be made less vulnerable to the residual stresses by reducing its width, i.e. by moving the second barrier closer to the first minimum.

In fact these modifications exactly correspond to the differences between the energy landscape of the L-selectin/PSGL-1 and that of biotin/avidin mapped by Merkel *et al.* [25] (reproduced in Figure 3C and 3D, respectively). This latter energy landscape is perfectly consistent with the known ability of the pair avidin/biotin to mediate firm adhesion under flow.

**1.2.3 Adhesions whose lifetime increases under a mechanical stress**

The L-selectin/PSGL-1 adhesion belongs to a class of bonds whose life-time is increased by a mechanical stress, even though this phenomenon is normally limited to a given range of forces, usually below 100 pN. This counterintuitive behaviour was first suggested almost 20 years ago on the basis of theoretical considerations [34], but while other peculiar phenomena were experimentally noticed before, like the actual increase of the binding probability with decrasing interaction times [35], it was demonstrated experimentally to exist in biology only in 2003 [36]. These bonds are usually called 'catch bonds', whereas the bonds showing the classic monotonic decrease of life-time with force increasing are called 'slip bonds'.

Catch bonds have been so far reported for carbohydrate-binding adhesion proteins like human P- and L-selectin [36, 37] and bacterial FimH [38], and for the myosin-actin motor protein interaction [39]. They are suspected also to be present in integrins [40]. The existence of catch bonds explains two long-standing problems in the biophysics of biological adhesion. The first is the irreversibility of specific biological adhesion: that is, the work required to peel off a unit area of adhesion is larger than the energy released from forming a unit area of adhesion [41, 42]. The second is the flow-enhanced adhesion, observed both in blood cells [43-48] and in bacterial adhesion [49-53]. In both cases, at forces higher than those promoting the catch bond behaviour, the firmness of the adhesion under flow is also sustained by 'classical' but finely tuned energy landscapes, as found by Fritz et al. [35] and Bustanji et al. [31]. In the case of blood cells, catch bonds have been suggested to have evolved to control platelets and leukocytes velocity under different flow conditions and to prevent spontaneous aggregation in 'stagnating' conditions. The discovery of catch bonds is another clear-cut proof of the importance, when studying biological adhesion, of thinking in terms of the reshaping of the energy landscapes by the forces developed in the environment where those adhesions take place.

Structural explanations that can support this paradigm also in the paradoxical case of catch bonds are just beginning to emerge. Thomas et al. [38] proposed, on the basis of physical considerations, an intriguing allosteric model for the catch bond behaviour of the FimH-mannose interaction. In this model, the deformation of FimH by a force switches the energy landscape of the FimH-mannose interaction to one with a higher life-time than the native one. Mechanical force in this case acts as an allosteric mediator that optimizes the mannose binding site of FimH.

In the case of P and L-selectins, the latest proposed mechanism [54, 55] is even more subtle. In this case, a small force could act by opening a flexible hinge region far from the selectin binding site. This opening changes the overall geometry of the selectin/ligand complex under force, tilting the interface to align with the force direction, thus allowing the ligand and the selectin to slide. In this conformation, when the main selectin/ligand bond is broken by the action of force, selectin slips on the ligand and can rebind it to a secondary site. This new transient interaction can also increase the overall binding lifetime by keeping the main binding site and the ligand close, thus giving the original bond the time to rebuild (see Figure 4A).

Catch bonds long escaped the single molecule community. A number of theoretical models for catch bonds were nevertheless proposed, even before their discovery (reviewed in [56]), but the concept was generally dismissed due to its counterintuitive nature. Previous experiments on selectins, for example, were analysed using classic slip bond models, and failed to detect possible hints of catch bond behaviour [56]. Lou and Zhu [55], following in turn the model of Barsegov and Thirumalai [57] in their detailed analysis of the 'sliding-rebinding' model for the L-selectin / PSGL-1 interaction, remark that the inability of the classic models of force-induced dissociation of correctly predicting catch bond behaviour, is due to their assumption of a single dissociation pathway. The energy landscape of the interaction is thus usually simplified as a 1-D one, where the only significant coordinate is the distance between the interacting surfaces. If at least another reaction coordinate, and therefore multiple dissociation pathways, enter into the description, the distortion of the energy landscape induced by the external mechanical force can influence not only the shape of a single dissociation pathway, but the actual route followed by the system on the potential energy surface. This, in turn, can affect the overall lifetime of the interaction (see Figure 4B). It is natural to think that biological systems have took advantage of this intrinsic complexity by evolving not only a finely tuned shape of a single pathway, like that mapped by Fritz et.al. [35], but possibly a choice of the dissociation pathways, according to the conditions.

The existence of catch bonds thus opens our eyes on the intrinsic multidimensionality of the energy landscapes of the mechanical dissociation of biochemical interactions. They also teach us that a meaningful description of this multidimensionality must be based on the unique structural features of the system itself.

## 2. THE TRANSMISSION OF A MECHANICAL STRESS BY A PROTEIN CHAIN.

The transmission and transduction of mechanical stresses within a cell has been reviewed very recently by Viola Vogel [9]. Three other recent reviews have appeared on closely related topics. The first review is on the cell capability of sensing and responding not only to the forces exerted by the environment but also to its own geometry, by the same Vogel in collaboration with Michel Sheetz [8], and by Disher *et al*. [3]. The third review article is on the mechanosensation through ion channels, by Ching Kung [2]. In this section, we look at those topics from the point of view of the single molecule methodologies. Single molecule methodologies based either on AFM or on Optical Tweezers are now making it possible not only to evaluate and compare the mechanical stability of the folding of the proteins that sustain the transmission of the mechanical signals, but also to elucidate the determinants of their mechanical properties. When these methodologies are supported by molecular dynamics simulations [11, 58-60] the atomic resolution is reached in the characterization of the conformational transitions that rule the mechanical properties of the different protein molecules.

## 2.1 The mechanical resistance of a protein molecule under a ramp of force

The complex system of transmission of forces within the cell relies on physically coupled extra- and intracellular force-bearing networks of proteins [9]. The energy landscape of a binding event that can trigger a biochemical signal should be therefore shaped so as to be compatible with the mechanical stresses delivered on it by this network. There is a major difference in the structural design schemes of the force-bearing proteins outside and inside the cell. Proteins of the extra-cellular matrix (ECM), as well as those that play key roles in linking the trans-membrane integrins to the contractile cytoskeleton consist of multiple and often repeating modules strung together to form extensible multimodular proteins.

The folding of these extracellular modules is normally sustained by beta-sandwich motifs. A great challenge is now to deepen our understanding of the engineering principles upon which the mechanical properties of these networks are based.

### 2.1.1 The first mechanical experiments on biological polymers at a single molecule level

Since the mid-1990s, different kinds of experiments have been performed applying external forces to single molecules of structured biological polymers in order to explore the mechanisms and the regimes of their mechanical unfolding. These experiments allow us to map the folding energy landscapes of those polymers [29] and understand the vectorial nature of their mechanically-induced unfolding [61-63].

Historically double-stranded DNA has been the first molecule stretched by the group of Carlos Bustamante [64]. Afterwards, by optical tweezers, the elastic response of B-

DNA was characterized in detail [65-67]. In particular the presence of a mechanically induced transition has been unveiled by the presence of a plateau in the force-extension curves. As shown in figure 5, at approximately 65 pN the DNA length start to increase at constant force up to 1.7 times its original value. This transition has been first interpreted as an overstretching of the DNA in which, upon pulling, the molecule originally in the B form transforms in the S form. The latter being characterized by the unstacking of the base-pairs and the consequent double helix rearrangement [65, 66]. More recently, several works appeared addressing the transition to the mechanically induced DNA melting [68]. This interpretation is supported by thermodynamics arguments and by the binding affinity to intercalating molecules [69].

**2.1.2 The mechanical unfolding of a single protein molecule**

Soon after these single-molecule experiments performed on DNA, stretching experiments were also done on single molecule proteins, and the abovementioned theoretical models that were used to interpret force plots like that in Figure 5 were then also utilized in the case of the proteins. In 1997 the group of H. Gaub reported the mechanical unfolding of titin at a single molecule level[70]. This protein was chosen because of its multi-domain structure. It consists of many similar domains, sharing a common primary structure, folded in individual three- dimensional structures (the so-called Ig domains) like a pearl necklace. As shown in the scheme in Figure 6, when the C- and the N-terminal are pulled apart at constant velocity (velocity-clamp mode), the sequential step-wise unfolding of the individual domains produces a sequence of saw-tooth like peaks in the force curve plot. Each saw-tooth like peak corresponds to the unfolding of an individual domain.

Those experiments opened the way to address with unprecedented effectiveness questions like: Which is the mechanical design principle that explains why many force-bearing proteins are rich in beta-sheets whereas others have a high alfa-helix content? How are these motifs composed so as to achieve specific elasticity and mechanical properties?

The mechanical unfolding forces of many beta-structures, like those of titin [70, 71] or tenascin [72], have been monitored by the Atomic Force Microscopy (AFM)-based SMFS methodology. Those modules were found to unfold at forces in the range of 100-300 pN, at loading rates of the order of $10^{-5}$ N/s. The alpha helix domains unfold instead, in the same conditions, at forces almost one order of magnitude smaller [73-75]. Also barnase, a beta-sheet rich protein without any putative mechanical role *in vivo*, seems to unfold at low forces in the same conditions [76].

Furthermore proteins with secondary structures based on a two-stranded coiled coil, such as that of myosin II, were found to behave as a truly elastic structure that can be unfolded at equilibrium. In this case, the unfolding and the refolding curves are superimposed [77]. In the case of secondary structures where the polypeptide chain is folded into itself with re-entrant structures, like that of titin domains, the unfolding and the refolding follow different pathways. This means that they take place in non-equilibrium conditions [78]. The typical saw-tooth pattern (Figure 6A) of the unfolding is absent in the refolding force curves. A marked hysteresis results between the unfolding and the refolding curves, and the area framed by the curves is equal to the amount of energy locally wasted as heat (Figure 6B).

The myosin II coiled-coil, a RNA stem-loop, a DNA double-stranded helix or a DNA single-stranded hairpin can behave instead as truly elastic springs (the unfolding and refolding traces are superimposable) mostly because their structures are topologically very simple. As in many other contexts, simplicity means speed - under an external force, the simple folding of these systems can relax very fast and therefore the proteins yield and sustain structural transitions from the folded to the unfolded structure that take place under quasi-equilibrium conditions [33, 66, 67].

As the topology of a system becomes more complicated, and a macromolecular domain involves interactions between residues that are farther away along the primary sequence (such as for β-sheets with respect to α-helices, for instance), the number of possible inter-atomic interactions that can be settled within an unfolding–refolding cycle increases enormously with the chain length. The time required to sample all these possible interactions in order to fall into the optimal energy minima becomes longer [79]. Whenever the rate at which the force is being applied to the molecular system is faster than its slowest relaxation time, the process takes place in non-equilibrium conditions. This happens because the system is not allowed to sample all its possible configurations during its mechanical unfolding.

**2.1.3 The mechanical role of sacrificial structures**

The peculiar mechanical role that can be played by the unfolding of a biopolymer with repeated structures, including multimodular proteins, was first recognized by Paul Hansma *et al.* [80]. Their attention was focused on the incredible strength and toughness of certain natural materials [81-83], such as spider dragline silk, whose breakage energy per unit weight is two orders of magnitude greater than high tensile steel [84, 85], or abalone shell, with its very high fracture resistance, or the mysterious molecular origin of the toughness of the bones [86]. They draw a very enlightening analogy between the extension under a force of a multimodular protein (see Figure 6A) and the Greek myth of Sisyphus. Sisyphus was condemned by Zeus to push a heavy rock up a mountain but just before reaching the summit the rock would always slip out of his hands and roll back to the bottom. In such a fall, the rock would dissipate its potential energy into heat, and Sisyphus had to start all over again. The case of extending a modular protein is analogous. When one stretches away the two ends of this type of protein before the breaking point of the polypeptide chain (the 'summit' in the Sisyphus myth) is reached, a domain unfolds, and the energy stored in the protein is dissipated as heat [84, 87]. Then, the constantly applied force puts the protein again under an increasing stress, until the next domain breaks and so on. These domains were termed 'sacrificial' because they are actually sacrificed in order to release the force before it can break the polypeptide chain. Only when no folded sacrificial domains are left, only after that point, will the chain finally break under the external force.

By the same principle, in the case of muscles, multi-modular proteins not only transmit stresses but also protect the muscle structure from damage whenever extreme forces are acting on it. In particular, the unfolding of the proximal Ig domains of titin may serve as a buffer to protect cardiac sarcomeres from being damaged at forces exceeding the physiological range [74].

This peculiar mechanical behaviour is well depicted by the schematic comparison in Figure 7 of the force curve of a multimodular protein with that of a main-chain unfolded polymer. Whereas the latter, upon pulling, elongates according to a single entropic behaviour, the former undergoes the typical saw-tooth pattern we have discussed above and represented in Figure 6A. The excess area below the saw-tooth profile, compared to the single extension one, is the extra amount of energy that the multi-modular molecule is able to absorb before reaching the breaking point.

Whenever the force externally applied to a multimodular protein is relaxed before the breaking of its main chain, the domains or loops refold, ensuring the typical elastic response of such materials. This response can actually change with the number of unfolding-refolding cycles. In fact the apparent titin contour length at low forces (0-25 pN) was found to tend to increase after various stretching-release cycles [88]. The authors attributed this phenomenon to the release of electrostatic interactions in the unfolded segments of titin (like the PEVK segment). These native interactions are easily broken by low forces, but require a relatively long time to rebuild correctly. If not enough time is given to the molecule to relax before the next stretching cycle, the molecule will appear to 'wear out' along cycles.

Because of this peculiar mechanical behaviour, a molecular bridge containing sacrificial bonds like that provided by a multimodular protein can modulate and maintain over great extensions the mechanical conversation between two surfaces that are moving apart [89].

### 2.1.4 Multimodular constructs for single molecule force spectroscopy experiments

The pioneering experiments performed with protein cardiac titin have also opened the way to the design of a new class of protein constructs to be used in force spectroscopy experiments. Several groups have conceived new multi-modular constructs consisting in repetitions of the same protein module or groups of protein modules. The presence of identical copies of a protein fold in the same multimodular construct allows us to combine the advantages illustrated above with increasing statistics of the unfolding events [57, 59, 71, 90]. In some works, different proteins were engineered in a defined order to introduce handles and internal gauges for the interpretation of the force-curve plots [71, 91].

The approach of introducing handles and internal gauges, similarly to the case of the multi-modular proteins, has been followed by a new approach developed by us to mechanically break single molecular interactions with the SMFS methodology. This has been made possible by synthesizing a branched polymer that makes it possible to obtain saw-tooth like force curves on breaking a sequence of identical interactions equally spaced along the polymer chain (to be published) as can be seen in Figure 8.

## 2.2 The mechanical resistance of a protein molecule under a complex waveform

The nano-manipulations we have mentioned so far were performed by, first, bridging the AFM tip to an underlying surface with the protein under investigation, and then by moving them apart, at constant velocity, by a piezoelectric actuator. On the other hand, force waveforms much more complex than this simple ramp of force can be met in biological systems.

**2.2.1 Examples of complex force waveforms**

The blood vessels in vertebrates display distinct microenvironments with characteristic mechanical stress waveforms, due to the interplay of the cardiac beat and the vessel geometry. These are pulsatory (Figure 9A) at the straight sections of the vessel wall and oscillatory (Figure 9B) at the reattachment points in the arterial branches. These two kinds of waveforms have been found to have differential effects both on the vessel endothelium and on tethered blood cells (reviewed in [92-94]). For example, oscillatory shear stress has been found to increase adhesion of monocytes to endothelial cells, while pulsating stress decreases it. Oscillatory stress also increased P-selectin and ICAM-1 gene expression in monocytes [95]. Pulsating and steady shear stress differentially affect the endothelial redox state [96] and the expression of endothelin and nitric oxide synthase in endothelial cells [97]. Cyclic strains also affect smooth muscle cell gene expression (see, for example, references [98] and [99]). In the case of membrane translocation the force required to unfold the proteins has a characteristic periodic waveform.

How are these large-scale (cell- and tissue-wide) effects built up at the level of the single protein molecules involved in these processes? Theoretical studies suggested that a periodic force perturbation on a protein can lead to stochastic resonance between the folded and unfolded states [100], and also can substantially affect the P-selectin/PSGL-1 bond lifetime [101]. Chtcheglova *et al.* [102], measured the spring constant of a complex antigen-antibody. This experiment simulates the response of a complex between two molecules, which for example are tethering two cells, to the thermally and mechanically driven stimuli induced by the fluctuations of the cells themselves. Several different biomolecules have then been studied by similar AFM-based techniques allowing both their characterization in terms of the viscoelastic properties and the discovery of new features in their mechanically induced transitions [103, 104]. However, a robust understanding of these phenomena at the single molecule level is still lacking. Single molecule techniques are just starting to tackle this level of complexity by applying cyclic and step-wise waveforms on single protein molecules, as shown in the following two examples.

The first is relative to the complex mechanism of protein import into mitochondria and the second to the unfolding-refolding activity of some molecular chaperones. Protein function in fact relies on their native three-dimensional folded conformation, but the cell may transiently require their unfolding [105]. This happens especially in two crucial cellular processes, the protein translocation across a membrane and the unfoldase activity of molecular chaperones necessary to recover the activity of misfolded proteins.

In the former case, proteins have to be transferred across a narrow pore, the diameter of which is too small to accept a folded domain; therefore the cellular system involved in such an activity has to induce the unfolding of the protein prior to translocation. The main factor responsible for this task has been identified in the 70 KDa molecular chaperone hsp70 (heat shock protein 70). Proteins that must be translocated across the membrane display a long portion, called presequence, that does not have an ordered structure and can thus reach the membrane and start entering the pore; hsp70 has been demonstrated to be able to bind to the presequence upon their passing through the membrane pore and actively pull them across it.

This activity of hsp70 requires the hydrolysis of ATP, the so-called ATPase activity, meaning that an active supply of energy is necessary. This is not surprising if one recalls that the work performed is not only the bare transport but also the active unfolding of the folded part of the protein to allow it to enter the channel. Hsp70 must apply a force on the protein to actively induce its unfolding and this activity is known as unfoldase. Several models have been employed to try explaining this activity without coming to common conclusions [106, 107].

Recently a work of De Los Rios *et al*. [108] has proposed an innovative description of this biological activity, providing also a description of the energy profiles and of the forces playing a role (see Figure 10). The model is called entropic pulling and is based on the simple assumption that upon binding of hsp70, the conformational freedom of the part of the protein that has crossed the membrane is reduced, thus increasing the free energy of the system because of the entropic change; to recover the free energy variation to the thermal energy values, the unfolded portion of the chain with the locked hsp70 has to be moved further from the membrane.

They demonstrated that a stretch of approximately 30 amino acids (aa) has to be pulled out from the pore to achieve the free energy recovery. The presence of hsp70 binding sites on average every 30 aa along the protein chains ensure that another hsp70 will then bind to the newly exposed site and act again as an active pulling machinery. The hydrolysis of ATP is necessary to change the affinity of hsp70 for the substrate protein and its cycle ensure the turnover of the hsp70 available for binding to the accessible binding sites.

According to this model thus the protein translocation machine will periodically apply a net force in the direction of the inner part of the membrane that will at the same time import and unfold the substrate. The force has been estimated to be in the range of 10-20 pN when the hsp is bound near the membrane (maximum value) then decreasing to the thermal level before the next pulling cycle.

A similar situation is faced by the cell when it has to deal with a misfolded protein; in this case the active unfolding machinery has to unravel the wrong structure in which the substrate protein is trapped to drive it along a refolding pathway. Two molecular chaperones are performing this job in bacteria, the previously described hsp70 and the chaperonin hsp60. The mechanism employed by hsp70 to unfold the substrate is the same previously described for the protein translocation. The binding and hydrolysis of ATP will induce its binding to the misfolded or aggregated proteins. Now the whole aggregate will play the role that before was that of the membrane. The entropic pulling action will apply a net force disrupting the misfolded protein.

The chaperonin hsp60 (in *E.coli* it is called GroEL) is a large protein complex formed by 14 subunits arranged in two seven-fold rings stacked one on top of another. This double ring structure is crucial for its activity - GroEL is in fact binding within its cavities unfolded or misfolded proteins. This is achieved by means of the hydrophonic surface exposed on the cavity surface that interacts with the abnormal amount of hydrophobic surfaces exposed by misfolded and unfolded proteins. Force-induced unfolding of the misfolded protein is then achieved because GroEL subunits, upon binding and hydrolysis of ATP, are undergoing a large conformational change that applies a mechanical load on the bound protein and, at the same time, converts the cavity surface from hydrophobic to hydrophilic, offering a local environment more prone to the correct folding of the sequestered protein (see Figure 11). The effect of this cycle is to put the misfolded protein on a higher free-energy point in the multidimensional energy landscape from which it can either reach the native conformation or be trapped in one of the competing basins of attraction. In the latter case it will be eventually re-bound to undergo another unfolding-refolding cycle till it reaches its native state. This mechanism is explained in details in a review published by George Lorimer and Devarajan Thirumalai in 2001 [109] and schematically described in Figure 13. The force applied both by the entropic pulling of hsp70 and by the power stroke of hsp60 have been estimated to be in the range of 10-20 pN and to follow periodic cycles [109].

Two experimental works recently appeared on the correlation between the mechanical stability measured by AFM and the unfolding rate [110, 111] (see Figure 12). A more recent theoretical one by Paci *et al.* [112] concluded that the difference between the geometry of the protein relative to the pore and the one probed by AFM experiments might hinder accurate predictions of the translocation rate from the mechanical properties. The physical principles governing the transformation of the metabolic energy in the active work performed by the molecular motors has been thoroughly reviewed by A. Vologodskii in a recent work [113].

These processes underline the importance of simulating, by single molecule techniques, ramps of forces such as those experienced by the proteins in the cell during the unfoldase activity necessary for the translocation and re-activation of misfolded and aggregated proteins. An experiment can be designed in which a misfolded protein is subject to different periodic ramps of forces, such as a periodic repetition of the entropic pulling originating from the free energy profile represented in Figure 10 thus monitoring the possible refolding under different ramp shapes. The possibility of mixing periods of constant forces, where the main cantilever could mimic a bound hsp70 subject to the thermal motion, to stretching at different speed, such as that applied by the GroEL subunits upon conformational change, could shed light into the different roles of the molecular chaperones and on their combined activity.

**2.2.2 Monitoring unfolding-refolding equilibria under a constant force**

Recently, thanks to the steady improvement of the quality of instrument technology, a new class of experiments has become possible - manipulations of single molecules at constant force. In 2001, Julio Fernandez and co-workers reported for the first time the construction of an instrument that allowed them to unfold titin, either keeping the applied force constant at a set value, in the so called force-clamp mode, or to increase the force linearly over time, in the force-ramp mode [114]. This was an important step towards the possibility of simulating in force spectroscopy experiments the complex force waveforms described above. This force-clamp approach has opened to AFM a way towards unfolding–refolding experiments under reversible conditions also on proteins with complex topologies. Those experiments were previously accessible only to set ups with optical tweezers, such as those used by Liphard *et al*. on RNA [33].

In these experiments, the pulling force can be maintained at a chosen value by continuously re-adjusting the probe–surface distance, and thus the length of the bridging molecule, through a feedback loop. Any sudden increase of the contour length of the protein, caused by a module unfolding, triggers a readjustment of the position of the AFM probe, the protein is thus immediately extended to an increased length, keeping the pulling force constant.

Because of the cantilever drift and the force feedback noise (about 50 pN), the experiments were actually done under only approximately constant force conditions. In spite of these instrumental limitations, the authors recorded the constant-force elongations of single molecules consisting of a sequence of titin I27 modules. Instead of the saw-tooth pattern recorded in the velocity-clamp mode (see Figure 6), a series of stepped increments in length was recorded, due to the unfolding of the different modules.

Furthermore, the possibility of quenching the force at any chosen value allowed observation of the refolding of these multimodular proteins under different mechanical loads [115] (see Figure 13). The time taken to fold was shown to be dependent on the contour length of the unfolded protein and the stretching force applied during folding. The folding collapse was marked by large fluctuations in the end-to-end length of the protein, but these fluctuations vanished upon the final folding contraction. Afterwards, the force was raised again in order to control the extent of the refolding.[115]

The interpretation of these results was done in the light of a polymer coil-to-globule transition, where the chain starts from an almost fully extended state and then collapse in a compact form as it would do a polymer placed in a poor solvent. In any case the recorded refolding trajectories are presently not clear enough to understand each detail of the folding of a complex object such as a multimodular protein construct. Further improvement in the experimental set up will make it possible to explore the refolding process in much more details. This result can be achieved through the employment of smaller force probe[116, 117] in order to strongly decreasing the noise level or through faster feedback controls.

Recent theoretical considerations have shown that protein refolding in this kind of force-quenching experimental set up may not be directly comparable to temperature-quenching protein refolding: the initial state[17] and also the reduced degrees of freedom[118] affect the preferred refolding pathways.

A conceptually similar experiment, but based on a totally different setup, was performed in Bustamante's laboratory and allowed following the unfolding and refolding of a single globular protein. In this design the polypeptide was not introduced in a multimodular construct but was tethered to the optical-tweezers-based sensor by mean of two long DNA molecules used as handles. The forces explored by this technique are much smaller than those in the AFM range, thus a finer tuning of the unfolding refolding process is possible. The *Escherichia coli* ribonuclease H (RNase H) was shown to unfold in a two-state manner and refold through an intermediate that correlates with the transient molten-globule like intermediate observed by bulk techniques. The unusual mechanical compliance of the intermediate was investigated: in fact it unfolds at forces substantially lower than those of the native state. The accuracy of their experimental setup allowed the hopping of the molecule between the unfolded and intermediate states to be observed in real time, and the energy landscape of the folding of this protein to be mapped with an unprecedented resolution.

### 2.2.3 Unfolding-refolding cycles

The force-clamp cycles of unfolding under a constant force and of refolding after the release of the force whenever reported in term of energy landscapes lead to diagrams like that in Figure 14. These cycles are represented by cyclic rides between iso-force energy curves. The release of the force at the end of an unfolding step induces the system to shift back onto a zero-force landscape (or onto a low-force landscape), and to refold thermally. The transitions connecting the two curves might be drawn as vertical segments if the system does not have time to re-equilibrate during the rapid onset or release of the applied force.

The cycle consisting in the application of an external force favouring the unfolding and the subsequent release, leading to the refolding, is well described by the two isoforce profiles marked in Figure 14. This way of representing the cyclic mechanical catalysis of processes provides not only a clear representation of experiments such as that reported in Figure 13, but allows a deeper insight in a biologically relevant process such as the previously described GroEL unfoldase activity. This molecular machine in fact to unfold the substrate and then refold, it applies a force that is subsequently released. The cyclic path of Figure 14 should actually be implemented in order to fit the GroEL case, because the applied force is not held constant but probably changes in intensity during its application (as during constant speed experiments). Upon the conformational change the protein inside the cavity is instead experiencing a force close to zero that remains constant, as in the force-release step in Figure 14. The multiple layers of force-tilted potential energy profiles are crossed in the case of pulling at constant velocity (velocity-clamp mode). The change of the applied force during the pulling or retraction makes the system shift amongst different iso-force profiles: the crossing path depends on the molecular system and the experimental conditions, such as the cantilever stiffness and the pulling speed.

## 2.3 The transmission of a mechanical stress brought about by a network of proteins

After having seen how a mechanical stress can be transmitted by a single protein, we must take into account that the cellular pipelines of transmission of the mechanical signals are in most cases based on force-bearing protein networks.

This implies that the stress acting over time, determining whether a module can pass an energy barrier and eventually unravel, is not controlled solely by its mechanical stability but depends also on the mechanical connections within the overall protein networks.

This problem was addressed by Hansma *et al*. [119] on studying the force curves recorded on stretching capture-silk threads. The non-linear increase of the elongation of these threads with the force does not fit the WLC model (see section 2.1.1) A purely exponential force-elongation behaviour was found instead (see Figure 15A). This behaviour was ascribed to their molecular network, that is composed of cross-linked chains. Therefore when capture-silk threads are stretched a collection of interconnected springs is pulled, and, as shown in Figure 15B, an exponential force-elongation behaviour can result from the combination of the elastic properties of the different springs within the network.

Those force curves revealed other mechanical details. Rupture peaks, typical of the multimodular proteins of about 60 pN, are superimposed onto the aforementioned exponential behaviour. The increases in force preceding those rupture peaks, and for intervals of 20-100 nm between two of them, were often linear. This is the typical behaviour of a Hookian spring The mechanistic origin of this property was not explained in this case. Recently, another example of a Hookean spring, but on a completely different sample, has been reported by Marszalek *et al*. [120] working with ankyrin repeats.

The complex mechanics of proteins networks is a topic that needs more comprehensive studies, starting from careful mechanical analyses of the individual components of the network and of the vectorial composition of their contributions.

# 3. CHEMISTRY CONTROLS MECHANICS: MECHANOCHEMICAL SWITCHES IN THE TRANSMISSION AND TRANSDUCTION OF A MECHANICAL STRESS

We have seen how the transmission and transduction of a stress into a biochemical signal is ruled by the overall energy landscape that controls the spread of the mechanical signals, the requested conformational changes at binding sites, and the dynamics of the resulting adhesion interactions that trigger the signal activation. We have also seen the role of specific structural motifs of the proteins in sustaining those processes. In this context we must also take into account that chemical functions present in the same proteins can make those energy landscapes liable to be modified by many environmental conditions, like pH, redox enzymes, ligand binding, etc. This can lead to processes much more complex and multifaceted than those described in the previous sections. This potential increase of complexity can be associated with more sophisticated controls of biochemical processes, as normally takes place in biology.

A mechanical stretching alone has several drawbacks when it is considered as a trigger of a biochemical signalling. For instance, the control of the extent of the stretching of a protein that corresponds to the stabilization of an intermediate can barely be tuned just on a mechanical basis. The generation and the transmission of a mechanical stress along a protein force-bearing network might lead to either a large variation or fluctuations (see section 2.2.1) in the intensity of the locally delivered force. We have recently proposed that a more robust tuning of a signalling process might be ensured by the coupling of a mechanical unfolding switch with a redox one based on an intramolecular disulfide bond [121].

Disulfide bonds are pervasive in the structure of extracellular proteins. This feature is commonly thought to have been selected by evolution to sustain and protect the native conformation of proteins in the relatively harsh extracellular environment. However this static view of the function of disulfide bonds has been challenged since the mid-1990s by a steady increase of evidence for extracellular regulation due to disulfide cleavage. Examples include secreted proteins like thrombospondin-1 [122], von Willebrand factor [123, 124] and plasmin [125], and cell surface receptors like the CD-4 T-cell receptor [126], integrins [127, 128] and the HIV protein gp120 [129]. Extracellular redox regulation has also been suggested as essential in activation of G-protein coupled receptors [130]. This regulation is in turn triggered by the action of oxidoreductases of the protein disulfide isomerase superfamily, that have been shown to control the extracellular thiol/disulfide equilibrium [131-134]. Lymphocyte activation, angiogenesis and tumorigenesis are among the processes that enhance the extracellular reductase activity [135-137]. Evidence for more specific pathways, like the disulfide cleavage of plasmin catalysed by PGK, has been found [138].

We might therefore conclude that an extracellular disulfide bond might act not just as an inert structural feature, but also as a reversible and flexible switch that might regulate specific protein functions. Furthermore, given that forces and disulfide bonds coexist in the extracellular space, we might also expect that mechanical and redox regulation have evolved to face and complement each other.

## 3.1 The disulfide bond can act as a redox switch that controls the mechanical properties of proteins

The interplay between disulfide redox equilibria and mechanochemistry has been studied in a few natural multimodular proteins [11, 139-141] and in artificially engineered titin modules [142, 143]. These studies have shown that the presence of an intramolecular disulfide bond results in dramatic consequences for the mechanochemistry of a protein.

The probability of a covalent disulfide bond failing under the weak forces generated outside or inside a cell is very low. A topological loop, whenever it is 'locked' by a disulfide bond, is therefore protected against the action of an external mechanical force. This locking has three main implications. The first is the modulation of the total mechanical extensibility of a protein. This was shown first experimentally in the case of V-CAM [141] and, more thoroughly, in angiostatin [139] (see Figure 16). The second implication is that potentially active binding sites buried in the interior of the loop enclosed by the disulfide bond cannot be activated by the external force (see below, section 3.1.1). Third, we might also imagine a case in which the main resistance point of the protein happens to lie inside a loop enclosed by the disulfide bond - the mechanical properties of the protein might drastically change according the redox condition met by that bond. Actually, to our knowledge, no evidence of this kind of modulation of the mechanical properties of a protein has yet been reported. Both in the case of human angiostatin [11] and of artificially engineered I27 titin modules [142] containing disulfide bonds, the main barriers to mechanical unfolding have been found to be almost independent from the reduction state. In the case of V-CAM mechanical unfolding, studied by Carl *et al.* [141], the position of the main unfolding barrier is unchanged, but its height is found to be slightly higher in the reduced state.

### 3.1.1 Disulfide bonds can add a fast and specific regulative control on top of the mechanical modulation.

Regulative mechanisms triggered by the mechanical deformation of a protein fold can meet a finer tuning when this deformation is controlled by the redox modulation of disulfide bonds. In fact, being able to sequester specific protein loops reversibly from the action of a mechanical force, a redox modulation should be expected to be able to bring specificity and plasticity to the response of a biological system to a force. In the absence of disulfide bonds, cryptic sites buried in the fold of a module can be potentially exposed each time a strong enough tensile stress is applied to the protein (see Figure 1). If the cryptic site is enclosed in a loop defined by a disulfide bond, the site can be mechanically exposed only after the disulfide bond has been reduced (see Figure 17, left). The same paradigm holds in the case of two binding sites spaced by a loop locked by a disulfide bond. In this case, if the synergic binding to both sites requires a stretching force to adjust their relative stereochemistry (see Figure 1), the unlocking of the disulfide bond is in turn required to get that concerted binding event to take place (see Figure 17, right).

An example of this kind of hierarchical control of protein mechanics by a redox chemistry has been recently proposed for human angiostatin by Grandi *et al*. [11]. Angiostatin has a strong antiangiogenic activity based on the inhibition of endothelial cell proliferation and migration. It is therefore acting in a mechanically active environment. It is a multimodular protein composed of five independently folded kringle modules (labelled K1-K5) that also individually show different extents of antiangiogenic activities (see [11] and references therein). Each module features a characteristic triple-loop topology due to three internal disulfide bonds. The SMFS methodology proved that redox and mechanical conditions that mimick those met by this protein on the surface of tumour cells populate a protein metastable intermediate of the K4 domain that exposes two cryptic polypeptide fragments that were hidden in the native structure. These fragments are highly active against cell migration and angiogenesis. The same mechano-redox conditions lead the K2 and K3 domains to populate a structure that sustains their concerted binding to the F1-ATPase of endothelial plasma membrane (a known angiostatin receptor). On this basis it was proposed by us that the active structure of angiostatin is not the native one, as commonly considered so far, but that partially reduced and partially unfolded whose characterization was made possible by the SMFS methodology [11]. The proposed mechanochemical pathway that activates angiostatin against angiogenesis is reported in Figure 18.

The mechanochemical model therein proposed can apply to other kringle fragments of multimodular proteins, like prothrombin, apolipoprotein (a) and hepatocyte growth factor [144, 145]. Moreover it can apply also to the main components of the basement membrane like thrombospondin-1, laminin or perlecan that, like ANG, are composed of independent modules containing internal disulfide bonds [146-148]. This mechanism can be of general importance in the extracellular matrix. It is in fact remarkable that all the known catalytically active cryptic sites in fibronectin have been found on the disulfide-stabilized FnI and FnII modules (see [9] page 477). For example, serine protease catalytic sites have been localized inside disulfide-closed loops of FnI modules [149, 150]. It was proved that these sites are exposed by a proteolitical pathway. On the other hand we should expect that it is entirely possible that a combination of disulfide reduction and forced unfolding can achieve the same result, with the benefit of the reversibility.

**3.1.2 The hierarchy of redox and mechanical regulation**

In the scheme in Figure 18 and the above discussion, a definite hierarchy between redox chemistry and mechanical regulation was assumed. The redox switch has been depicted as the upstream regulator that controls the activation of a downstream mechanical switch [121]. Evidence is emerging however that the also reverse is possible – that is, mechanics can control chemistry. Specifically, redox regulation in the extracellular space can be influenced by mechanical forces. Bhasin *etal*.[140] have observed that the kinetics of reduction of a disulfide bond buried in modules of the adhesion molecule VCAM-1 can be greatly enhanced by mechanical unfolding of the surrounding protein fold. Steered molecular dynamics simulations indeed showed that the buried disulfide bond becomes exposed in the very first steps of the mechanical unravelling of the modules. This makes it possible for the reducing agent in solution to attack the buried disulfide. In turn, the opening of the disulfide bond allows for further unravelling of the module. In the case of VCAM-1, therefore, multiple redox and mechanical switches may be sequentially activated.

A recent work by Wiita *et al*. [141] proved that an external mechanical force can directly alter the reduction kinetics of the covalent disulfide bond, independently from its accessibility. In this work, a disulfide bond has been artificially engineered on I27 titin modules and then kept at a constant force by means of the single-molecule force clamp technique, in the presence of DTT. The reduction rate of the bond has been measured to be exponentially dependent on the applied force.

The picture emerging from these results is that mechanical forces can significantly increase the reactivity of disulfide bonds in the extracellular space, both directly and indirectly. This adds to the already mounting evidence that extracellular disulfide bonds are dynamic switches whose importance in extracellular signalling is just beginning to be appreciated. On the other hand the coupling of redox and mechanical switches does not necessarily implies a distict hierarchy between them. Their intertwining can move across multiple layers of complexity that can potentially have an even broader influence on biological processes than expected so far.

# CONCLUSION

Crucial details of the pervading role played by the world of forces acting in most of the cellular functions, outside and inside the cell, are increasingly disclosed by our new capabilities of measuring at the single molecule level the kinetics and the lifetime of a molecular interaction, of measuring directly the forces in processes in which stresses and strains are developed along the reaction coordinate, and of applying an external force to alter the extent and the fate of these processes. Our present understanding of the subtle intertwining between the mechanics of those forces and the chemistry of many cellular processes is enough to alert us to the need for constantly asking two main questions in the course of our biochemical studies. The first is: Are we dealing with systems where mechanical stresses might be developed? If this is the case, the force must also be introduced to describe the thermodynamics and kinetics of the biochemical reactions taking place in those systems, and the possibility of bringing into play single molecule force spectroscopy experiments should be considered. The second question, directly connected to the previous one, is:

Are we sure that the proteins involved in those processes act in their native form, and not with a structure that has been mechanochemically induced by their environment? These two questions imply an important change in our conceptual and practical approaches to biochemical studies.

**ACKNOWLEDGEMENTS**
This work was supported by MIUR-FIRB RBNE03PX83/001; MIUR-FIRB Progetto NG-lab (G.U. 29/07/05 n.175); PRIN 2007, EU FP6-STREP program NMP4-CT-2004-013775 NUCAN, EUROCORES-SONS program BIONICS, FIRB Progetto RBNE03PX83–006

LEGENDS

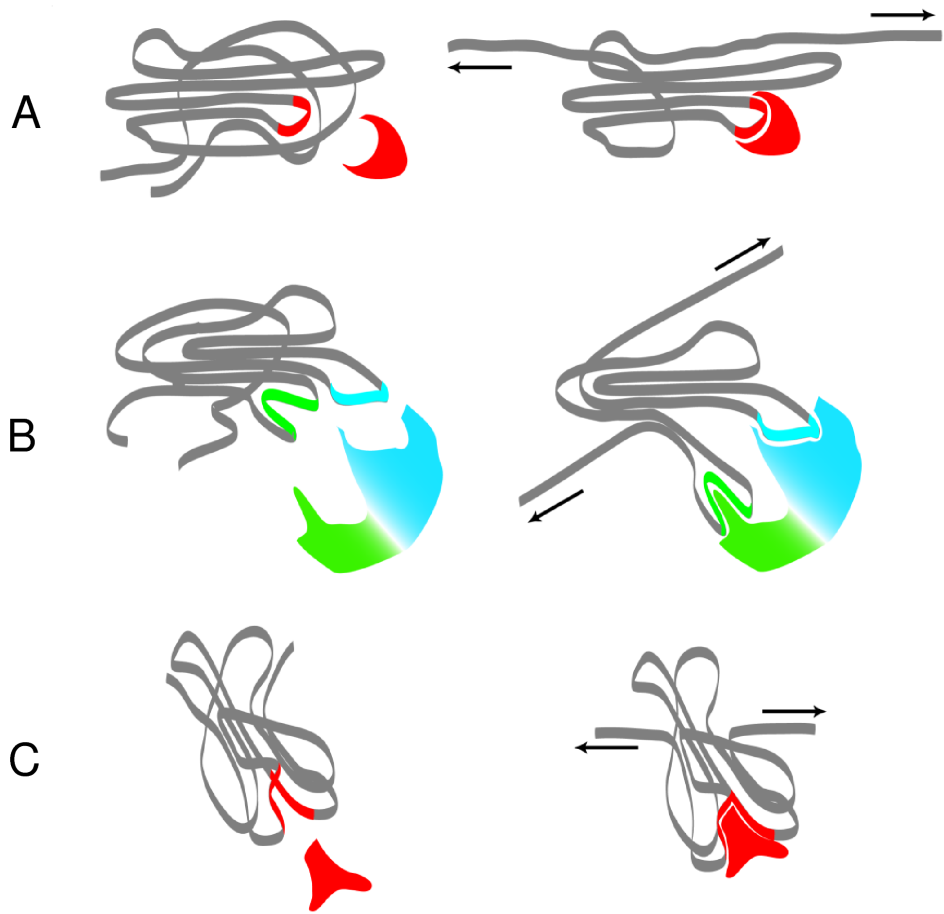

**Figure 1**
The conversion of a mechanical tensions into biochemical signals by force induced conformational transitions. A) By exposure of cryptic sites: a mechanical force can stretch a protein into a conformation in which a binding site, previously buried in the folded native structure (left), is exposed and thus enabled to bind a receptor (right) B) By changing the distance of binding sites: two sites that are not at a correct distance to bind cooperatively to a receptor can be brought by a force to a relative distance that makes their concerted binding possible. C) By changing the shape of the binding site: the force can modify the geometry of a potential binding site in order to make its binding to a ligand possible. Figure reproduced from [121].

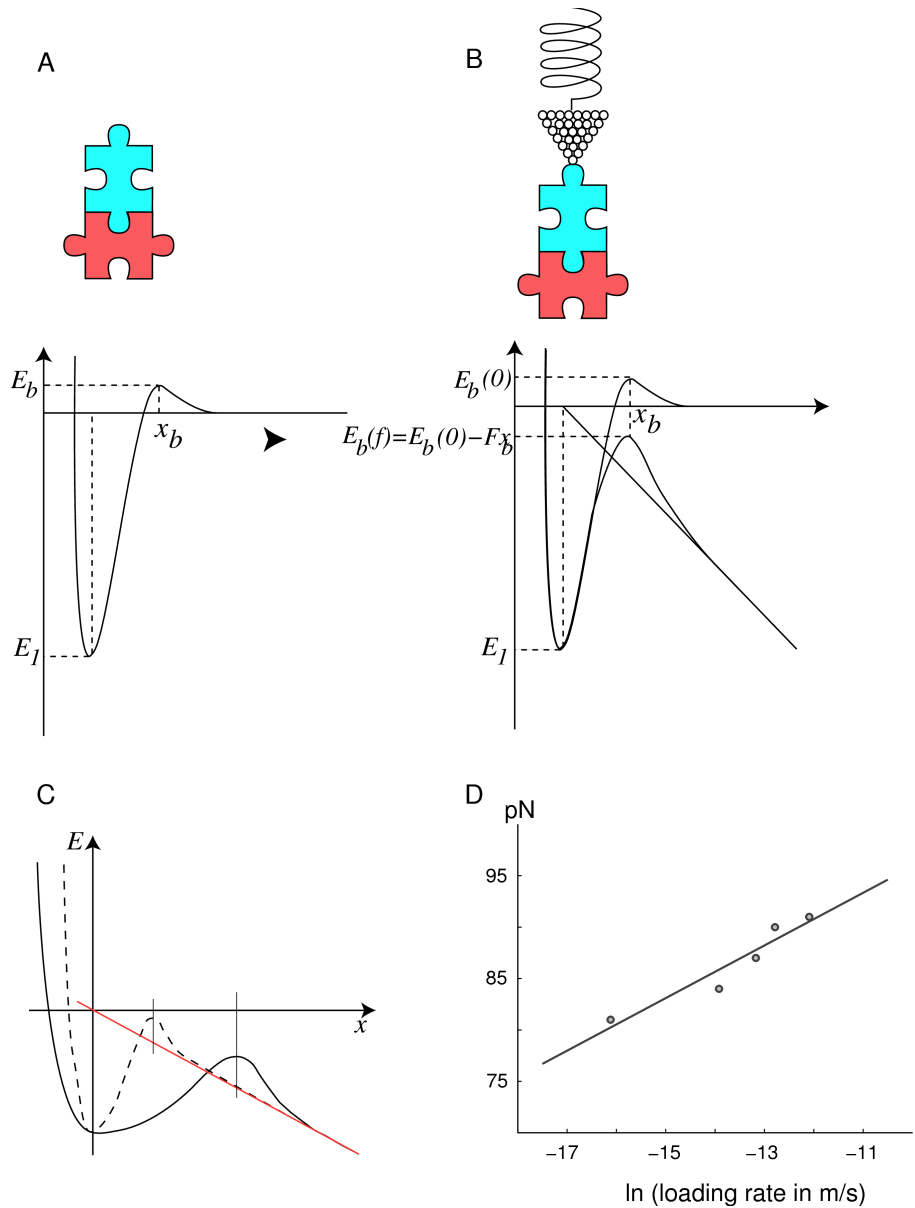

**Figure 2**
The energy landscape of a binding interaction is tilted by the application of an external stretching force. A) Energy landscape at zero force of a pair of interacting molecules. A barrier whose height is $E_b$ is located at $X_b$ along the reaction coordinate -that is, the distance between the molecules. B) When a force is applied the free energy profile is tilted downwards by an amount equal to the mechanical work $Fdx$ that is done on the system (represented by the straight tilted line), where F is the stretching force and $dx$ is the elongation along the force vector. C) The extent of the reduction of the height of a barrier is greater the further is the location $xb$ from the equilibrium configuration. D) A force vs loading rate logarithmic plot of the most probable unfolding force of angiostatin kringle modules upon a mechanical stress (from [11]). The plot shows a linear dependence between the force the loading rate $kv$ (where $k$ is the elastic constant of the cantilever and $v$ the pulling speed).

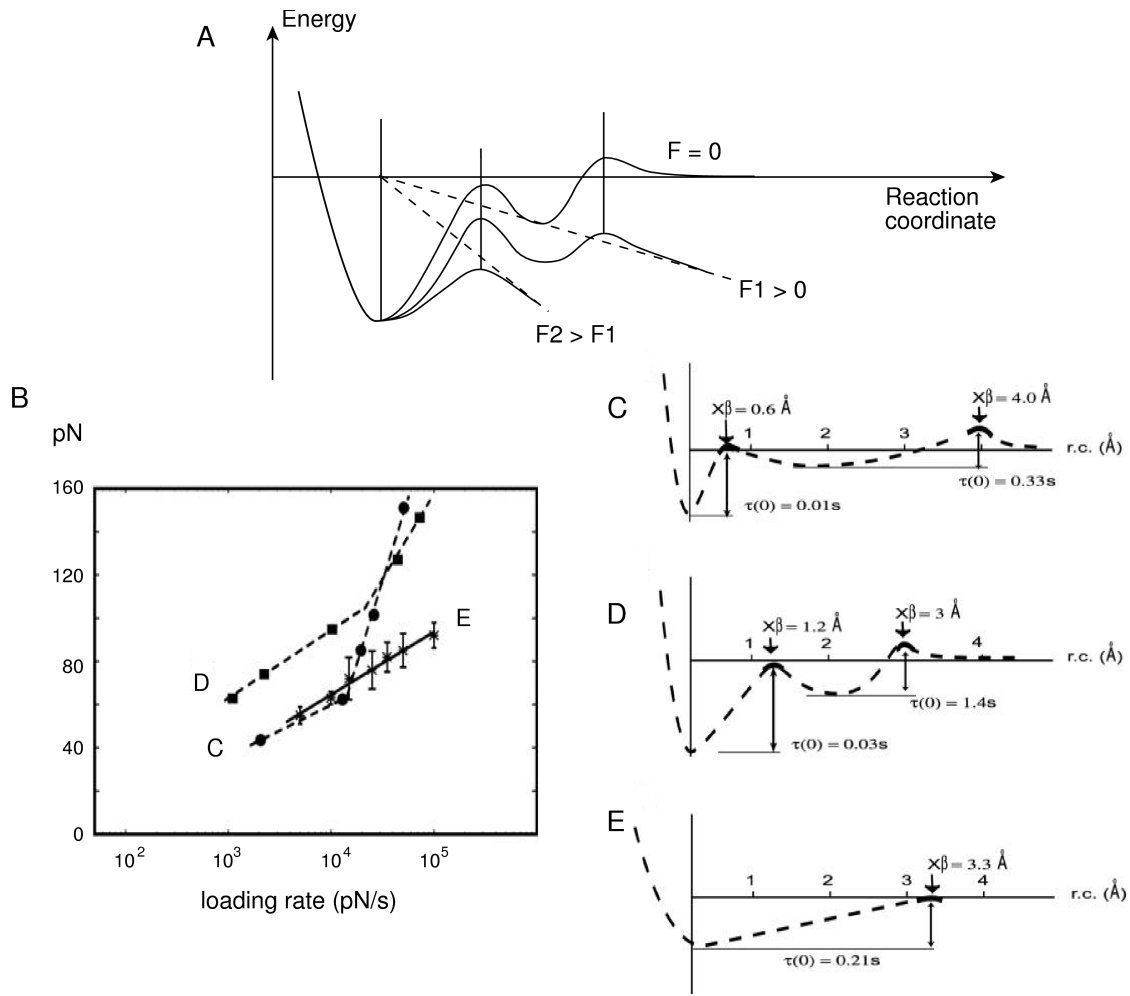

**Figure 3**
A) Energy landscape with two main barriers separating the bound state from the dissociated one. Let's suppose that at zero force the inner barrier (next to the equilibrium configuration) be lower than the outer barrier. The latter is therefore the rate-limiting barrier at low forces. Upon stretching the bond the outer barrier is tilted downwards more than the inner barrier (see Eq.6), until the inner barrier becomes the rate limiting one. B) Force vs loading rate logarithmic plots of the force-induced dissociation of three adhesive interactions: C: L-selectin/PSGL-1 D:avidin/biotin E:fibronectin/bacterial fibronectin-adhesin The abrupt changes in the linear slopes were interpreted classically as the fingerprint of a multiple-barrier energy landscape, thus indicating the loading rates at which a hidden, inner barrier has became the kinetically dominant barrier, overcoming the previous one. From the analysis of the linear dependency for each slope it was possible to determine the position of each barrier on the energy landscape. C, D, E) Sketches of the energy landscapes whose barrier positions were extrapolated from the plots in B). Panels B,C,D,E partially edited from [31].

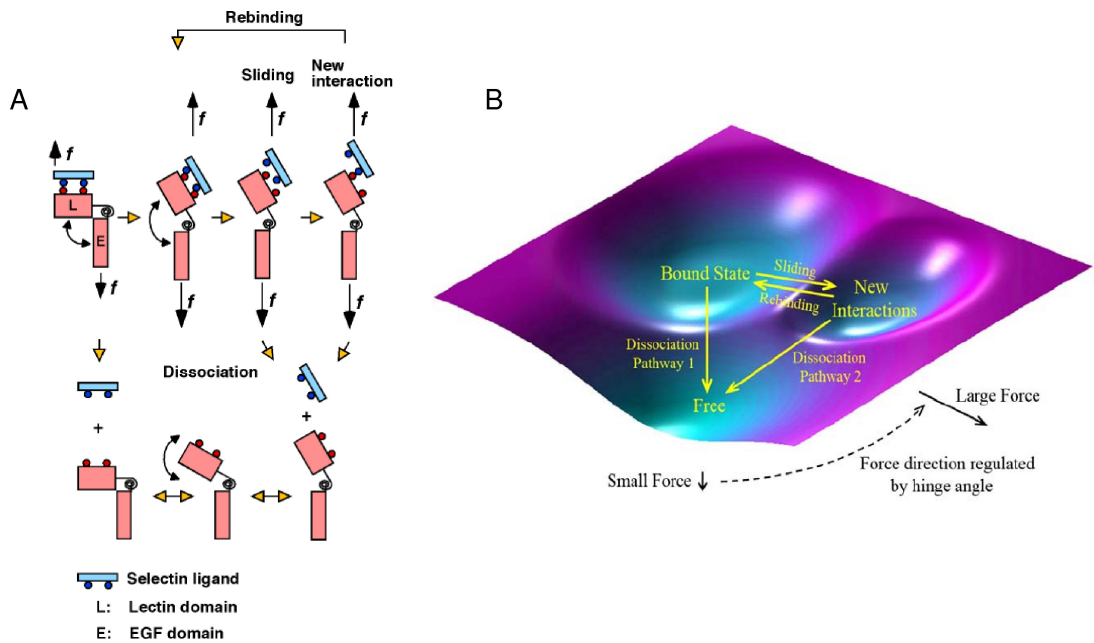

**Figure 4**
How the sliding-rebinding model may explain the catch-bond behaviour of the L-selectin/PSGL-1 interaction. A) Sketch of the sliding-rebinding model. The lectin and EGF domains of L-selectin are bound by a flexibile hinge. At forces that are too weak to bend the flexible hinge (top left), the lectin domain stays perpendicular to the force vector. In this conformation force-induced dissociation is favoured (fast pathway). With increasing forces the hinge can bend and the force vector can become parallel to the bimolecular interface (center to right). In this other conformation, the ligand is forced to slide across the interface. During sliding new interactions can form and keep the ligand and the lectin domain bound or the ligand can slip back to the original interaction site (slow pathway). Eventually the complex dissociates, but the overall dissociation kinetics will be slower than at low forces. From [54], with permission. An atomic detail description of this sliding-rebinding mechanism of L-selectin/PSGL-1 has been obtained by SMD simulations [55]. B) Sketch of an energy surface that models the sliding-rebinding mechanism by taking into account multiple dissociation routes that can be chosen by the system. The molecular complex can dissociate from the bound state following pathway 1 but, if new interactions take place, it can be instead trapped into another energy well. From here, the system can fall back into the bound state or dissociate following pathway 2. Increase of the force magnitude initially brings the system from the direction of pathway 1 towards the direction of the sliding-rebinding pathway thanks to the opening of the interdomain angle. The overall intermolecular complex lifetime is thus increased. Further increase of force lowers all energy barriers until dissociation is always accelerated, regardless of sliding and rebinding, marking the transition from catch bond to slip bond behaviour. From [55], with permission.

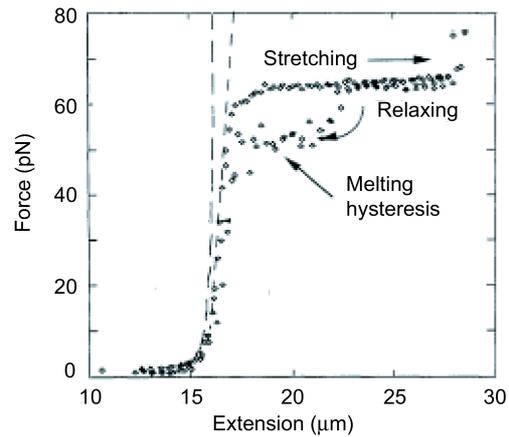

**Figure 5**
The overstretching transition of B-DNA, as detected by single molecule force spectroscopy. In the plot adapted from Smith et al. [66] the initial extension of the B-DNA, fitted by wormlike chain model (— —), is followed by the transition plateau in the stretching curve. In the relaxation curve an hysteresis appears due to the double strand melting. The plot with short dashes (- -) is the real behaviour of the DNA chain and it is displayed to underline the difference with the inextensible wormlike chain model due to presence of the elastic deformation.

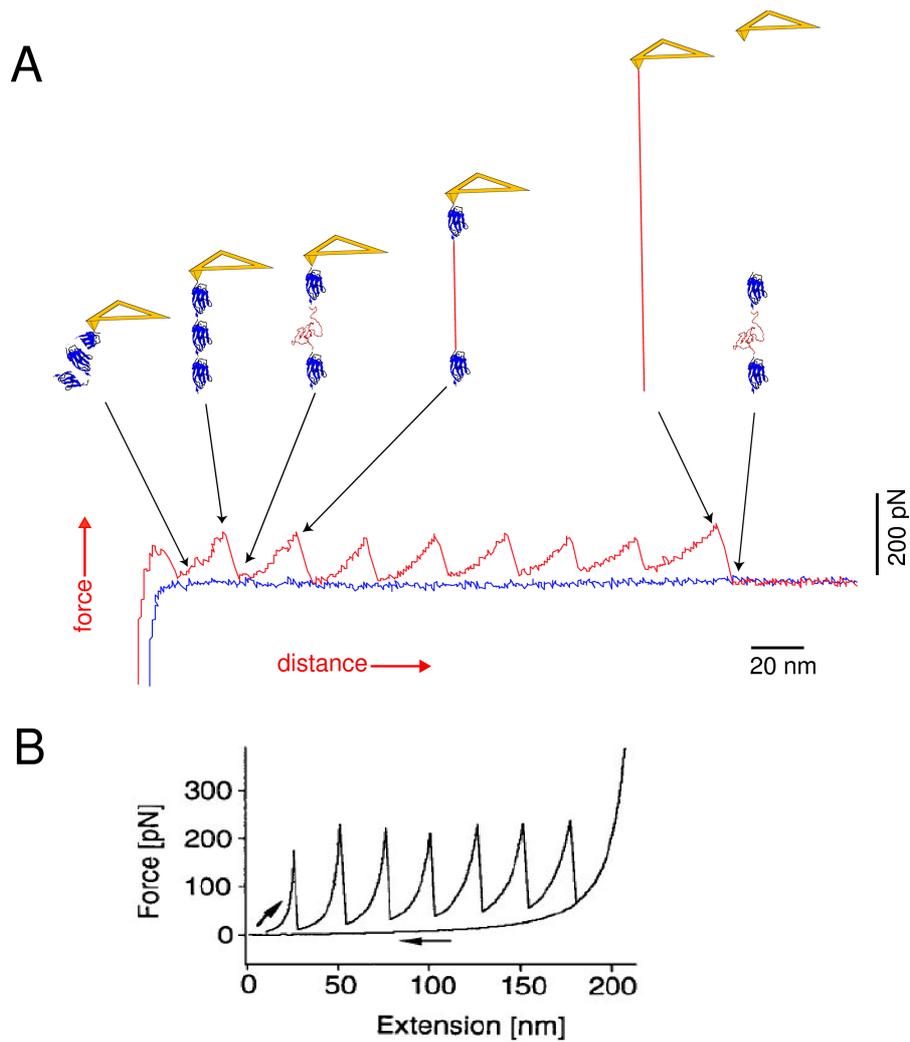

**Figure 6**
A)Fig. 3. Unfolding multimodular proteins by AFM single molecule force spectroscopy. In a typical AFM force spectroscopy experiment the cantilever tip approaches the surface, pushes on it and then retracts. The typical force curve corresponding to a multimodular protein unfolding experiment shows a saw tooth profile, where each dominant force peak represents the unfolding of one domain. Immediately after the unfolding of each domain, the tension is released and the force drops down. As the extension keeps increasing again, the unfolded module begins to unravel, behaving like a random coil chain, and the force starts rising up again due to the entropic elasticity of the chain. The increased tension applied on the still folded modules leads suddenly to the unfolding of another domain. B) Unfolding-refolding traces of a multimodular protein, showing a marked hysteresis. Reproduced with permission from [20].

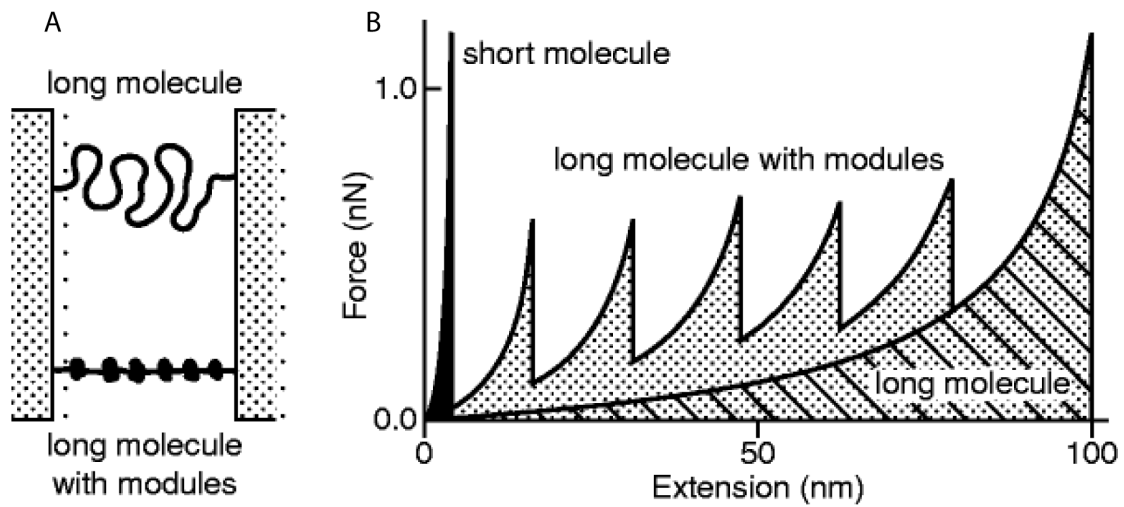

**Figure 7**
Force-extension curves for three different kinds of molecules. A short molecule (solid curve) resists pulling up to a high force before it breaks at small extensions. The energy required to break this short molecule (area under the curve) is small. A long molecule (hatched curve) behaves like an entropic spring and yields to the pulling force up to large extensions. The energy to break the long molecule is larger than that for the small molecule, but the forces at low extensions are small. On the contrary, a long molecule that is compacted into domains (the stippled plus the hatched curve) that are held together with intermediate-strength bonds resists pulling already at small extensions. Before the molecule's backbone can break, modules unfold. The sequential unfolding of the modules allows stretching up to large extensions, requiring in turn signicant energy. Therefore, a long molecule that is compacted into domains that are held together with bonds of intermediate strength combines both high (tensile) strength and high toughness. Figure and caption edited from [80].

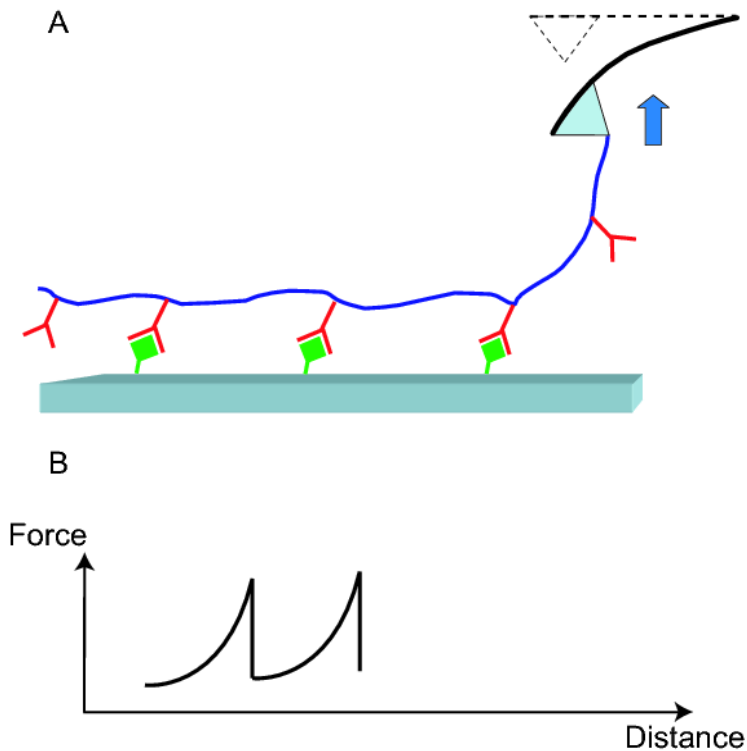

**Figure 8**
A) A long Poly-ethylen-glycole (PEG) based branched polymer (blue) synthesized to expose side chains (red) displaying the desired reactive groups at fixed distances along the main chain. This comb-like polymer, once adsorbed onto a substrate presenting the appropriate binding sites, forms a certain number of specific bonds between its side chains and the groups present on the substrate. To measure the rupture forces of these bonds, it will be enough to pull the branched polymer away form the surface using the AFM probe. The force-extension profile of the desorption of the polymer from the surface will in turn display unbinding events at the positions corresponding to the specific anchoring points. The distance between these events is determined by the separation of the side chains along the polymer. B) The force curves measured will have a saw-tooth profile mimicking the pattern of the unfolding of multimodular proteins.

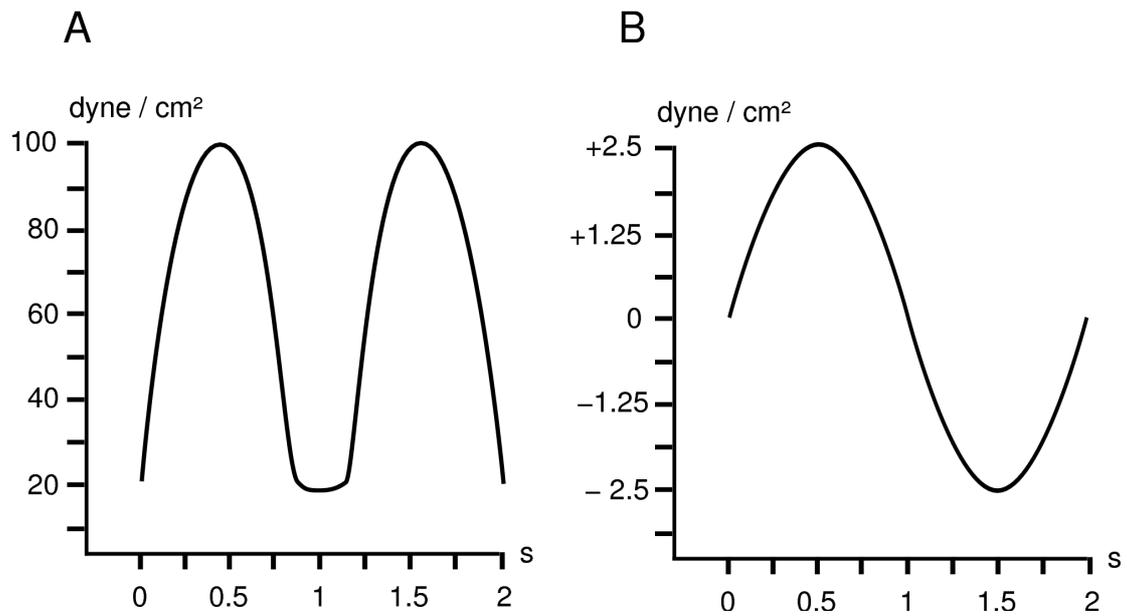

**Figure 9**
Examples of the shear stress waveforms that are applied on blood and endothelial cells by the combination of blood flow and heart pulse in the human circulatory system. A) pulsatile flow, typical of the straight portion of the arterial walls. B) oscillatory flow, typical of the outer wall of arterial bifurcations at the reattachment points. Adapted from data in [95].

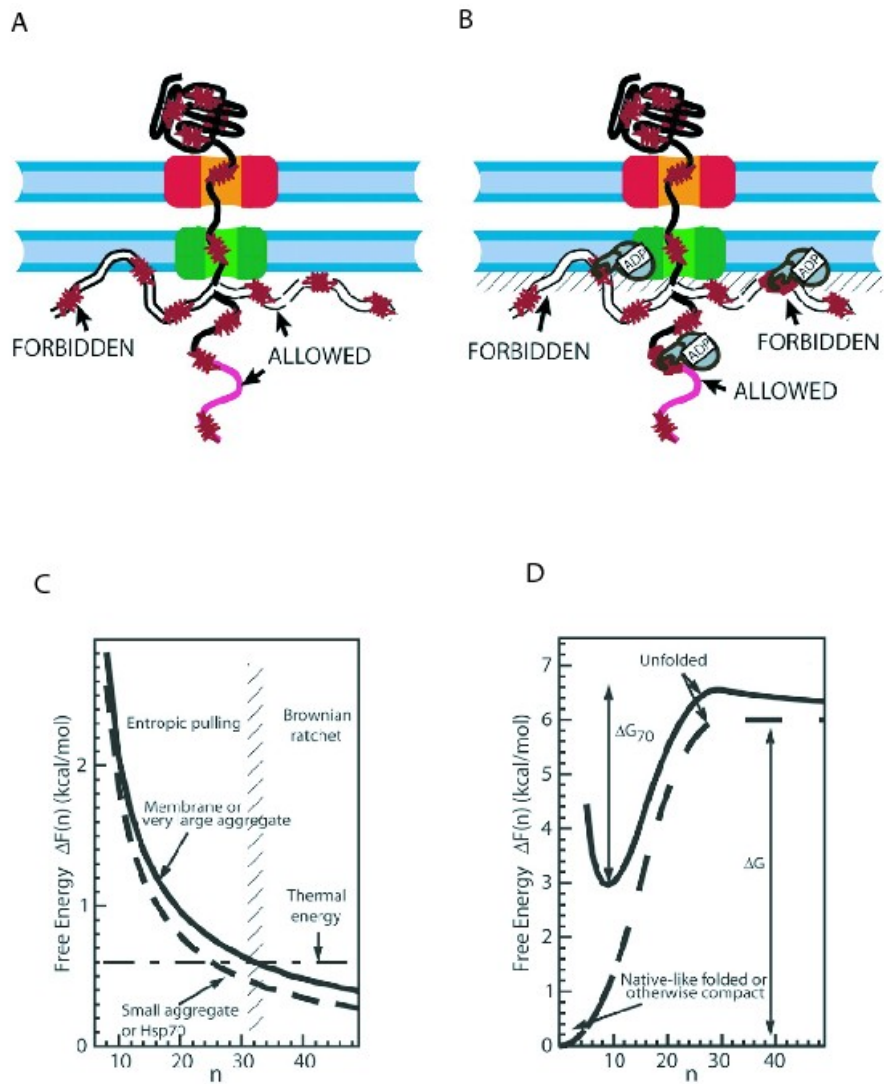

**Figure 10**
Schematic representation of the excluded volume effect leading to the mechanism of the entropic pulling. A) The presequence of the protein to be imported enjoys a high conformational freedom. B) The binding of hsp70 forbids to the polypeptide chain the conformations corresponding to the shaded region near the membrane, thus applying a net force of entropic origin on the protein. C) Free-energy profiles of polypeptide translocation and aggregate unfolding by action of Hsp70 calculated as a function of a parameter *n* corresponding to the number of imported residues in the mitochondrial matrix. The horizontal dot-dashed line is the energy associated with thermal fluctuations and the vertical shaded region at residues 31–33 separates a region (to the left) where entropic pulling prevails from a region (to the right) where entropic pulling is least effective D) Free-energy profile of unfolding for a native-like, misfolded or otherwise compact translocating protein in the absence (dashed line) or presence (solid line) of a polypeptide-bound Hsp70 chaperone on the matrix side of the membrane. Figure edited from [108]

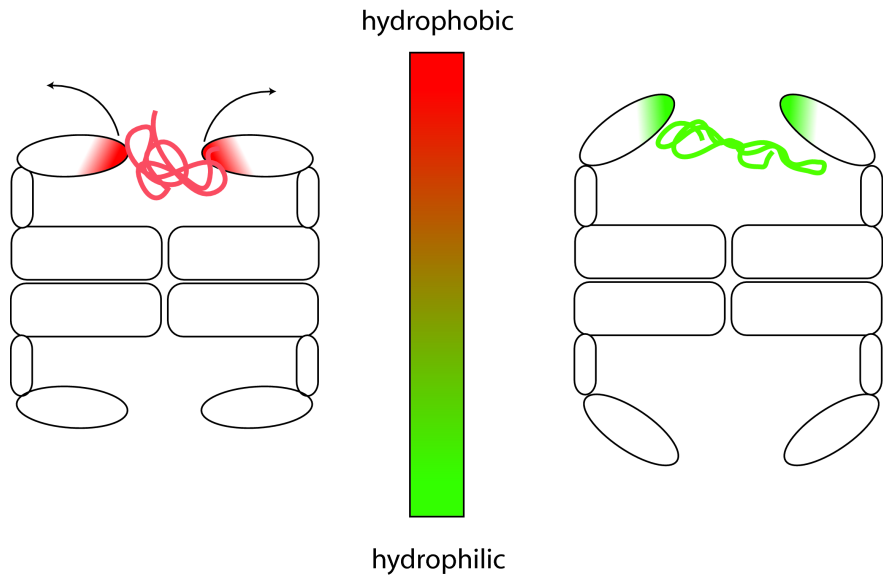

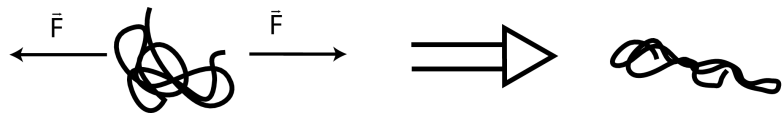

**Figure 11**
Schematic of the mechanism of activity of GroEL as reported in the current literature. The hydrophobic amino acids present onto the GroEL's cavity interact with the large amount of hydrophobic surfaces exposed by the unfolded/misfolded polypeptide thus leading to a strong binding. The large ATP driven conformational change of the subunits of one ring leads to both the application of a force on the substrate protein (as represented in the lower panel) and to the release of the polypeptide within the cavity by the switching of the cavity surface from hydrophobic to hydrophilic.

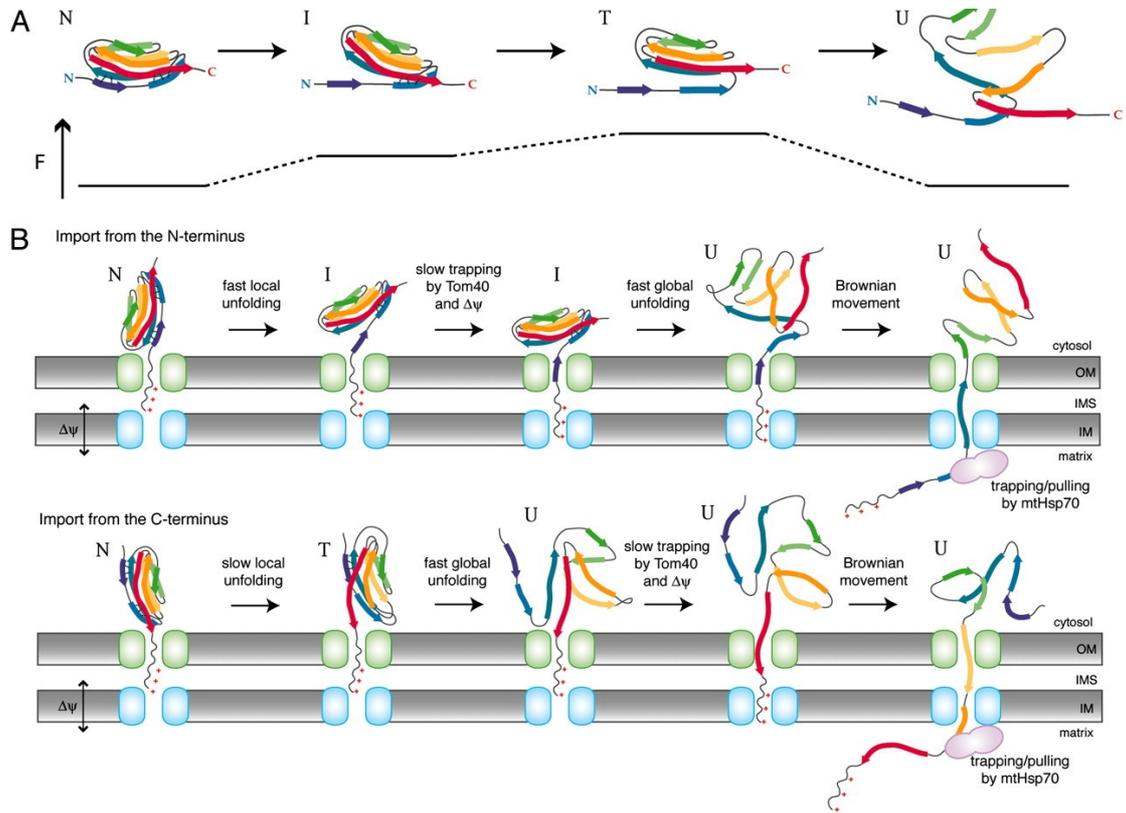

**Figure 12**
Mechanical unfolding of an I27 domain under three different conditions: A) stretched by an external probe, such as the AFM tip, between the C- and the N-terminal; B) upon translocation through a membrane pore pulling from the N-terminus or the C-terminus. In the work from which this figure has been extracted, Sato et al. compared the effect that some mutations, that alter the I27 mechanical stability, have on the import rate through a pore. The alteration of the I27 mechanical stability was previously characterized by AFM mechanical unfolding by the group of Julio Fernandez.
On the basis of this scheme it is possible to understand why the geometry of the domain stuck against the rim of the pore might have an influence in decreasing the mechanical stability of the protein imported [110]. Due to the vectorial nature of the pulling force, the mechanical unfolding necessary to import the protein, may be helped or hindered according to the relative geometry of the pore and the portion of the protein still folded. This effect follows from a previously published paper by Brockwell and coworkers [61].

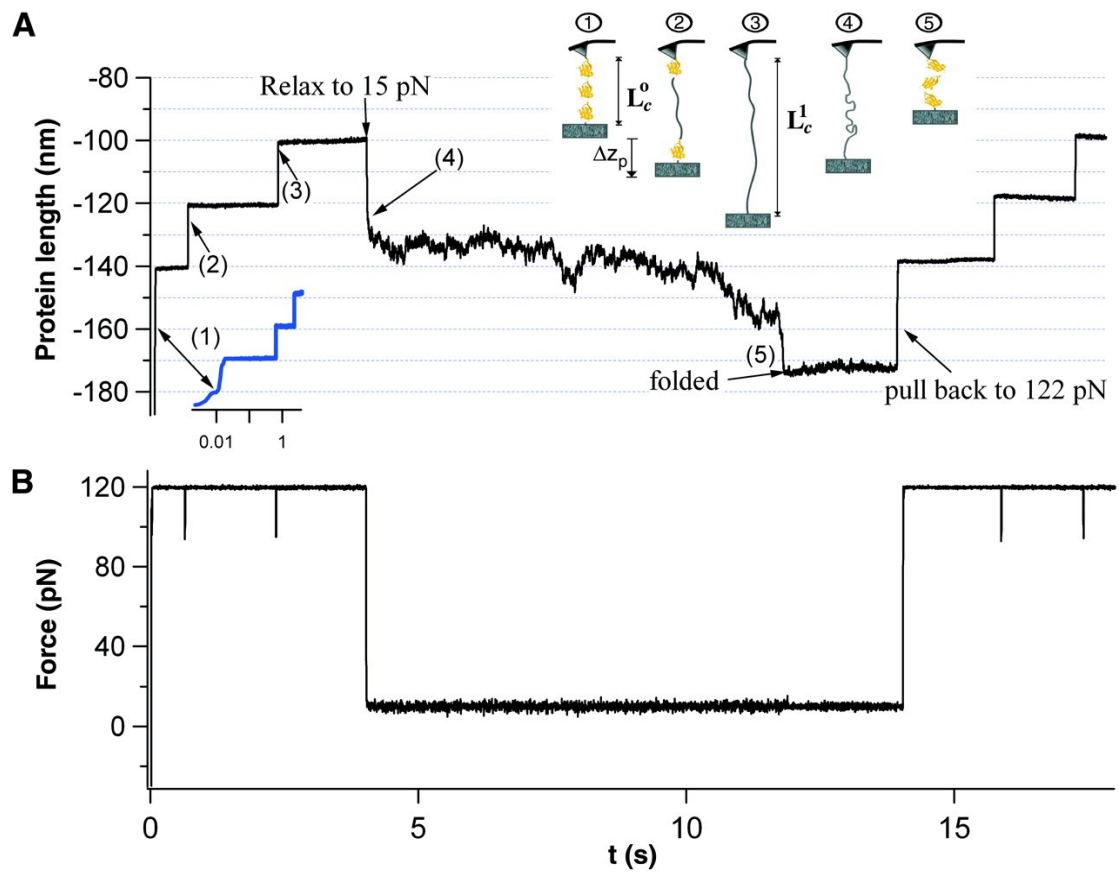

**Figure 13**
Forced unfolding, refolding and unfolding again of a ubiquitin polyprotein in force-step mode. A) Plot of the chain extension as a function of time, B) Corresponding plot of the applied force. The external applied force is first set at a high force (122 pN) to encourage the folded protein to unfold (steps 1-2-3 correspond to the unfolding of three ubiquitin domains) then released to a lower force (15 pN) to increase the likelihood of the refolding, and finally set to 122 pN again. If in this last phase unfolding signals appear again, this means that refolding was successful. It is possible to see different stages in the refolding trace: the fast elastic recoil of the unfolded polymer (4), and then a continuous but fluctuating collapse as the protein folds (from 4 to 5). At 14 s the polyprotein was again stretched back to 122 pN. A stepwise pattern analogous to that at the beginning is present, confirming that the polyprotein modules had indeed folded again. The initial steplike extension is the elastic stretching of the folded polyubiquitin.The inset line above the time scale shows the different refolding stages. Figure edited from [115]

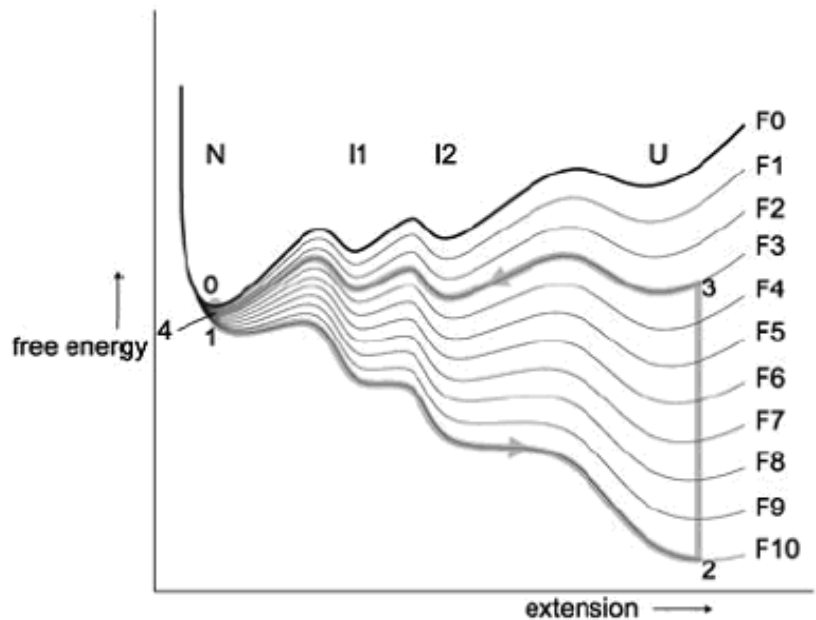

**Figure 14**
Energy landscapes along a force clamp experiment where the protein is initially subject to an external force that induce its unfolding and then to a smaller one which allows the refolding in its native form. This kind of representation takes its inspiration from the thermodynamic description of thermal cycles and allows a direct comparison between the energy dissipated and the mechanical work performed during the activity for instance of the GroEL unfoldase activity.
Looking in the details of the force profiles: the system at equilibrium at the point 0 on F0 is taken to the point 1 on F10 by the external force, then relax to the higher extension that corresponds to the new local minimum at that applied force. Afterwards, if the first barrier (now the highest) is thermally surmounted, the system unfolds to the unfolded state U. At the end of unfolding (point 2, the lowest energy point on F10), the external force is released to a lower value, thus making thermal refolding more probable. This transition is, again, supposed to be quicker than the relaxation time of the system (and is thus drawn vertical) with an end-point that resides on an energy profile where the unfolded state is less stable than the folded one; F3 might be the profile. In this case the folded (N) state is only slightly more stable than the intermediates, and energy barriers at similar heights separate the intermediates, which are at similar energies The thermal surmounting of the first barrier (from U to I2) practically allows the system to populate the intermediate states, with similar probabilities. Eventually, the system will fold to state N at point 4 on F3.
Figure edited from [116]

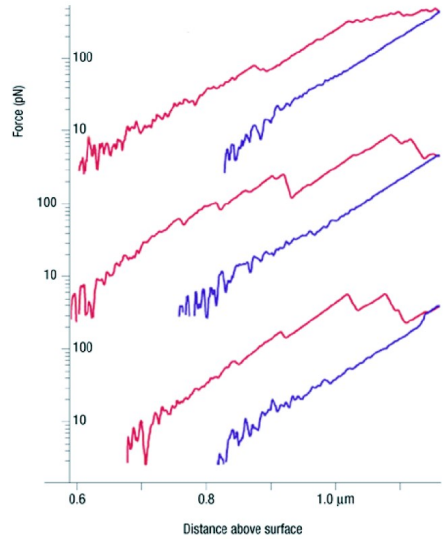

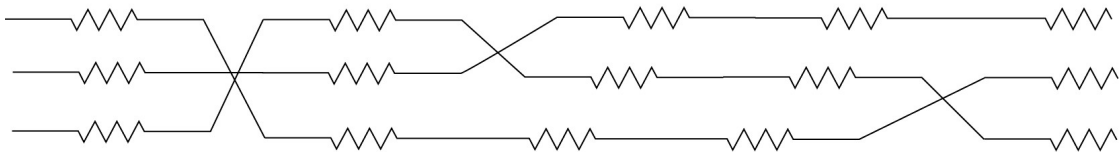

**Figure 15**
A) Exponential force vs extension profiles measured by pulling spider capture silk molecules with the AFM [119]. This behaviour has been addressed to the networked structure of the fibers, due to crosslinking between the molecules, as schematically sketched in B). See [119] for further details.

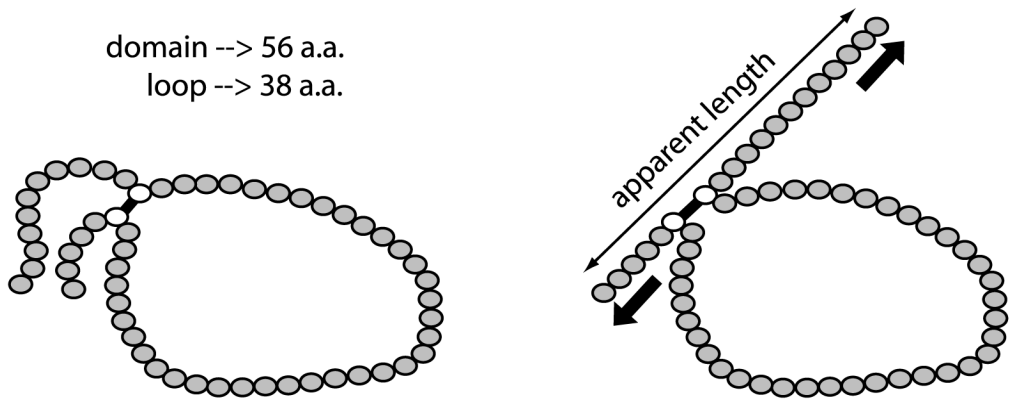

**Figure 16**
Disulfide bonds hide portions of a protein domain from an external force. A) Two dimensional sketch of a protein domain of length L containing an internal disulfide bond. The resulting topology of the domain allows to distinguish an inner loop of length $L_{loop}$ enclosed by the disulfide bond. B) The disulfide bond acts as a barrier to mechanical unfolding. The apparent contour length $L_{app}$ that will be directly accessible to force in the same protein sketched in A), is therefore given by $L_{app} = L - L_{loop} + SS_{bond}$, where $SS_{bond}$ is the length of the disulfide bond itself (this can be safely neglected). If $L_{loop}$ encloses a large fraction of total L, the presence or absence of a disulfide bond can effectively switch the protein between two radically different elastic states: a locked, inextensible one and an unlocked, compliant one. Figure and caption originally published by the authors in [121]

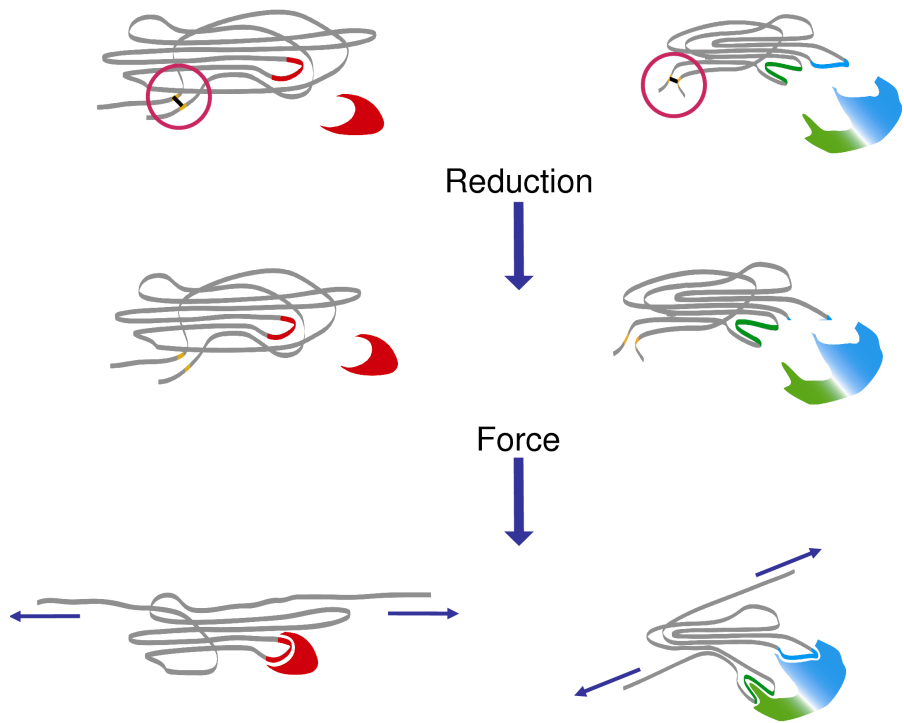

**Figure 17**
The coupling of cysteine redox and mechanical regulation. Left: If a cryptic site is enclosed in a loop defined by a disulfide bond, mechanical force alone is not able to break the constrain and free the site. Only after the reduction of the disulfide bond, the protein fold can be mechanically unraveled and the cryptic site revealed, thus allowing binding. Right: The same logic applies for two synergic binding sites that must be placed at the correct distance to simultaneously bind their target. Also in this case reduction of the disulfide bond makes it possible to deformate the protein fold as to allow the correct displacement of the synergic binding sites.

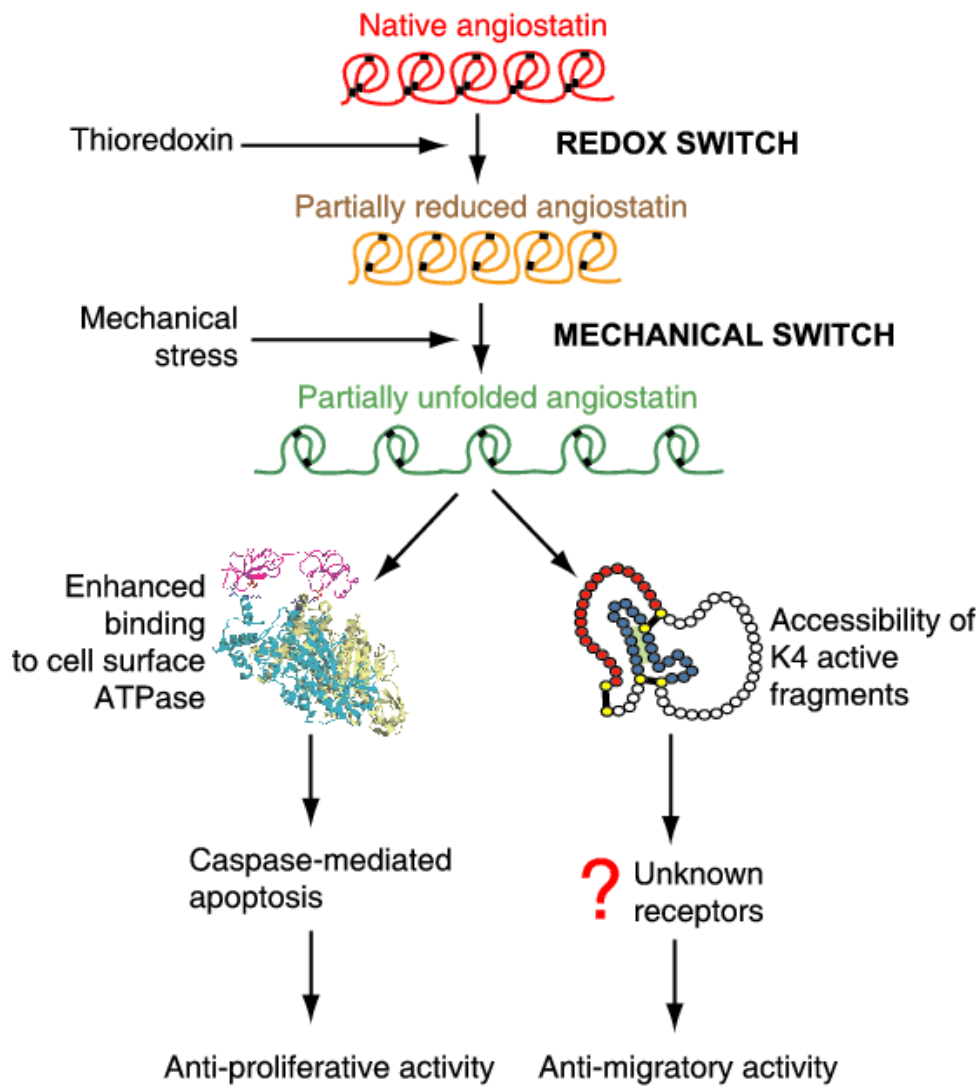

**Figure 18**
The mechano-redox regulation paradigm as proposed for human angiostatin by Grandi *et.al.*[11]. The reducing environment met by ANG on the surface of highly metastatic tumors is able to reduce the most external disulfide bonds of its kringle domains. This reduction creates the possibility that the mechanical stress that is constantly being developed in the ECM environment, where ANG is located, can mechanically unfold a topological loop of 40 amino acids. The resulting partially elongated, and partially unfolded structure of ANG can activate new binding capabilities and trigger new bio-chemical signals, as shown for the cases of a K2–K3 fragment and the K4 kringle domain. The hierarchical activation of the redox and the mechanical switches can lead i) to an enhanced binding of a K2–K3 fragment to ATPase and therefore to an increased antiproliferative activity of ANG; ii) to a more efficient exposure of the two segment chains that ensure, through the inter-action with still unknown receptors, antimigratory activities higher than that of the native K4 full domain. Figure and caption originally published by the authors in [11]